\documentclass[]{aa}
\usepackage{xspace}
\usepackage{graphicx}
\usepackage{natbib}
\usepackage{amsmath}
\usepackage{amssymb}
\usepackage{subfigure}
\usepackage{float}
\usepackage{deluxetable}

\newcommand{\nraoblurb}{The National Radio Astronomy Observatory is
a facility of the National Science Foundation operated under cooperative
agreement by Associated Universities, Inc.}
\newcommand{\hide}[1]{}

\newcommand{\micron}{\mbox{$\mu$m}}%

\newcommand{\gb}{\ensuremath{{\it b}}\xspace}
\newcommand{\absb}{\ensuremath{\vert\,\gb\,\vert}\xspace}

\newcommand{\cm}{\ensuremath{\,{\rm cm}}\xspace}

\newcommand{\K}{\ensuremath{\,{\rm K}}\xspace}

\newcommand{\ghz}{\ensuremath{\,{\rm GHz}}\xspace}

\newcommand{\degree}{\ensuremath{^\circ}\xspace}

\newcommand{\jy}{\ensuremath{\,{\rm Jy}}\xspace}
\newcommand{\jyb}{\ensuremath{{\rm \,Jy\,beam^{-1}}}\xspace}

\newcommand{\mjyb}{\ensuremath{{\rm \,mJy\,beam^{-1}}}\xspace}

\newcommand{\hi}{{\rm H\,{\footnotesize I}}\xspace}
\newcommand{\hii}{{\rm H\,{\footnotesize II}}\xspace}

\begin{document}

\title{Galactic Supernova Remnant Candidates Discovered by THOR}

\author{L.~D.~Anderson}

\author{L.~D.~Anderson\inst{\ref{wvu},}\inst{\ref{gbo},}\inst{\ref{center}}
  \and Y.~Wang\inst{\ref{mp_h}}
  \and S.~Bihr\inst{\ref{mp_h}}
  \and H.~Beuther\inst{\ref{mp_h}}
  \and F.~Bigiel\inst{\ref{u_h}}
  \and E.~Churchwell\inst{\ref{madison}}
  \and S.C.O.~Glover\inst{\ref{u_h}}
  \and Alyssa~A.~Goodman\inst{\ref{harvard}}
  \and Th.~Henning\inst{\ref{mp_h}}
  \and M.~Heyer\inst{\ref{umass}}  
  \and R.S.~Klessen\inst{\ref{u_h}}
  \and H.~Linz\inst{\ref{mp_h}}
  \and S.N.~Longmore\inst{\ref{liverpool}}
  \and K.M.~Menten\inst{\ref{bonn}}  
  \and J.~Ott\inst{\ref{nrao}}
  \and N.~Roy\inst{\ref{bangalore}}
  \and M.~Rugel\inst{\ref{mp_h}}
  \and J.D.~Soler\inst{\ref{mp_h}}
  \and J.M.~Stil\inst{\ref{calgary}}
  \and J.S.~Urquhart\inst{\ref{bonn},}\inst{\ref{kent}}
}
\institute{Department of Physics and Astronomy, West Virginia University, Morgantown WV 26506, USA \label{wvu}
\and Adjunct Astronomer at the Green Bank Observatory, P.O. Box 2, Green Bank WV 24944, USA \label{gbo}
\and Center for Gravitational Waves and Cosmology, West Virginia University, Chestnut Ridge Research Building, Morgantown, WV 26505, USA \label{center}
\and Max Planck Institute for Astronomy, K{\" o}nigstuhl 17, 69117, Heidelberg, Germany \label{mp_h}
\and Universit\"at Heidelberg, Zentrum f\"ur Astronomie, Institut f\"ur Theoretische Astrophysik, Albert-Ueberle-Str. 2, D-69120 Heidelberg, Germany \label{u_h}
\and Harvard-Smithsonian Center for Astrophysics, Cambridge, MA 02138, USA \label{harvard}
\and Department of Astronomy, University of Wisconsin-Madison, 475 N. Charter street, Madison, WI 53706 \label{madison}
\and Department of Astronomy, University of Massachusetts, Amherst, MA 01003-9305, USA \label{umass}
\and Astrophysics Research Institute, Liverpool John Moores University, 146 Brownlow Hill, Liverpool L3 5RF, UK \label{liverpool}
\and Max Planck Institute for Radioastronomy, Auf dem H\"ugel 69, 53121 Bonn, Germany \label{bonn}
\and National Radio Astronomy Observatory, P.O. Box O, 1003 Lopezville Road, Socorro, NM 87801, USA \label{nrao}
\and Department of Physics, Indian Institute of Science, Bangalore 560012, India \label{bangalore}
\and Department of Physics and Astronomy, University of Calgary, 2500 University Drive NW, Calgary AB, T2N 1N4, Canada \label{calgary}
\and School of Physical Sciences, University of Kent, Ingram Building, Canterbury, Kent CT2 7NH, UK \label{kent}
}
   
\date{Received/Accepted}

\begin{abstract}
  {There is a considerable deficiency in the number of known supernova
    remnants (SNRs) in the Galaxy compared to that expected.  This
    deficiency is thought to be caused by a lack of sensitive radio
    continuum data.  Searches for extended low-surface brightness
    radio sources may find new Galactic SNRs, but confusion with the
    much larger population of \hii\ regions makes identifying such
    features challenging.  SNRs can, however, be separated from
    \hii\ regions using their significantly lower mid-infrared (MIR)
    to radio continuum intensity ratios.}
  {Our goal is to find missing SNR candidates in the Galactic
    disk by locating extended radio continuum sources that lack
    MIR counterparts.}
{We use the combination of high-resolution 1-2\,\ghz\ continuum data
  from The HI, OH, Recombination line survey of the Milky Way (THOR)
  and lower-resolution VLA 1.4\,\ghz\ Galactic Plane Survey (VGPS)
  continuum data, together with MIR data from the \emph{Spitzer}
  GLIMPSE, \emph{Spitzer} MIPSGAL, and WISE surveys to identify SNR
  candidates.  To ensure that the candidates are not being confused
  with \hii\ regions, we exclude radio continuum sources from the WISE
  Catalog of Galactic \hii\ Regions, which contains all known and
  candidate \hii\ regions in the Galaxy.}
{We locate 76 new Galactic SNR candidates in the THOR and VGPS
  combined survey area of $67.4\degree > \ell > 17.5\degree$, $\absb
  \le 1.25\degree$ and measure the radio flux density for 52
  previously-known SNRs.  The candidate SNRs have a similar spatial
  distribution to the known SNRs, although we note a large number of
  new candidates near $\ell \simeq 30\degree$, the tangent point of
  the Scutum spiral arm. The candidates are on average smaller in
  angle compared to the known regions, $6.4\arcmin\pm 4.7\arcmin$
  versus $11.0\arcmin\pm 7.8\arcmin$, and have lower integrated flux
  densities.}
{The THOR survey shows that sensitive radio continuum data can
  discover a large number of SNR candidates, and that these candidates
  can be efficiently identified using the combination of radio and MIR
  data.  If the 76 candidates are confirmed as true SNRs, for example
  using radio polarization measurements or by deriving radio spectral
  indices, this would more than double the number of known Galactic
  SNRs in the survey area.  This large increase would still, however,
  leave a discrepancy between the known and expected SNR populations
  of about a factor of two.}
\end{abstract}

\keywords{\hii regions -- supernova remnants -- methods: aperture photometry -- radio continuum: ISM }

\authorrunning{Anderson et al.}
\titlerunning{THOR SNR Candidates}
\maketitle

\section{Introduction}
\label{sec:intro}
There is a severe discrepancy in the number of detected supernova
remnants (SNRs) in the Galaxy compared to that expected.  The most
authoritative recent compilation contains just 294 SNRs
\citep[][hereafter G14]{green14a}, but based on OB star counts, pulsar
birth rates, Fe abundances, and the SN rate in other Local Group
galaxies, there should be $\ga 1000$ \citep{li91, tammann94}.  These
estimates derive in part from studies of similar external galaxies,
scaled to the Milky Way based on its luminosity.  The discrepancy may
not be due to a true deficiency of Galactic SNRs, but rather may hint
at observational problems related to lack of sensitivity and confusion
in the Galactic plane \citep[e.g.,][hereafter B06]{brogan06}.

The Galactic supernova (SN) rate is an important parameter for
understanding the properties and dynamics of our Galaxy.  Most SN
arise from the core collapse of massive stars
\citep[cf.][]{tammann94}, and therefore the number of SNRs in the
Galaxy is tied to recent massive star formation activity.  SN inject
energy into the interstellar medium (ISM), driving molecular cloud
turbulence and galactic fountains out of the disk \citep{deavillez05,
  joung09, padoan16, girichidis16}.  This feedback can determine the disk scale
height and star formation properties of a galaxy \citep{ostriker10,
  ostriker11, faucher13}. The search for new Galactic SNRs is
therefore important for understanding the global properties of the
Milky Way.

SNRs are frequently identified at radio wavelengths.  According to the
G14 catalog, $\sim\,90$\% of known SNRs are detected and well-defined
in the radio regime, $\sim\,40\%$ detected in X-rays, and $\sim\,30\%$ in
the optical.  The radio emission is due to synchrotron radiation,
which dominates the Galaxy's low-frequency radio emission.  The
  most common radio morphology in the G14 catalog is that of a shell,
  or a partial shell.

Since many types of objects emit radio emission similar to that of
known SNRs, additional criteria are used to determine if a radio
continuum source is a true SNR.  These criteria are: 1)~the radio
spectrum of candidate has a negative spectral index (typically
$\sim\,-0.5$), 2)~the radio emission from the candidate is polarized,
3)~the candidate has associated X-ray or cosmic ray emission, and/or
4)~the candidate has a mid-infrared (MIR) to radio continuum flux
ratio much lower than that commonly found for thermally-emitting
plasmas.  The first two criteria can distinguish between thermal (flat
spectrum, unpolarized) and non-thermal (negative spectral index,
polarized) radio emission.  The third criterion is sensitive to
high-temperature ($\sim\,10^7\,\K$) plasma within SNRs that is rarely
detected in \hii\ regions.  The fourth criterion has characteristics
of the other three, in that it can also distinguish between thermal
and non-thermal emitters. The MIR emission from dust arises from the
interaction of the SN shock wave with the ISM during the initial
expansion phases \citep[e.g.,][]{douvion01}.  Other non-thermal radio
continuum sources such as active galactic nuclei can be excluded from
SNR searches due to their small angular sizes.

Many researchers have shown that SNRs are deficient in MIR emission
compared to \hii\ regions \citep[e.g.,][]{cohen01, pinheiro11}.  For
SNRs to produce MIR emission, they must be sufficiently dense to
produce collisional heating \citep{williams06}.  \citet{pinheiro11}
found that a typical 24\,\micron\ to 1.4\,\ghz\ flux density ratio for
SNRs is $\sim\,5$, although they found flux density ratios ranging from
0.5 to 10.  This low MIR to radio flux ratio holds even for young
regions like Cas~A, despite their strong MIR emission
\citep[see][]{rho08}.  Due to its powerful discriminatory power and
relative ease of use, the MIR to radio flux ratio is of most interest
here.

SNR candidates can be identified efficiently in radio continuum
surveys using their low MIR to radio continuum flux ratios.  While
there is some faint associated MIR emission detected for some SNRs
\citep{reach06, pinheiro11}, this emission is quite weak.  At radio
frequencies high enough that \hii\ regions are optically thin,
$\gtrsim\,1\,\ghz$, the MIR to radio flux ratio for SNRs is about 100
times lower than that of \hii\ regions.
\citet{helfand06}, hereafter H06, used the lack of MIR emission as one
criterion to identify 49 new SNR candidates in The Multi-Array
Galactic Plane Imaging Survey (MAGPIS) 20\,cm data.  B06 also used
this criterion to identify 35 SNR candidates in their VLA data.
Recently, \citet{green14b} used the anti-correlation between radio and
8\,\micron\ emission to identify 23 new SNR candidates from Molonglo
Galactic Plane Survey (MGPS) data.

These previous studies have first identified promising radio continuum
candidates, and then examined their 8.0\,\micron\ emission to
determine their classifications.  This method, however, has an
inherent bias toward objects that look like SNRs, i.e. shell-type
structures, at the expense of other possible SNR morphologies.  A
better method is to first identify all \hii\ regions from their MIR
morphologies and high MIR to radio continuum flux density ratios, and
to then locate radio continuum sources not associated with the
\hii\ regions.  This removes the confusion from \hii\ regions in the
Galactic plane, which is a major difficulty in new SNR identifications
given their potentially similar radio morphologies and the much higher
spatial density of \hii\ regions.
By excluding \hii\ regions, one can search for non-thermal emission
features without imposing any source morphology bias.

Here, we identify extended sources of emission in radio continuum data
from The \hi, OH, Recombination line survey of the Milky Way
\citep[THOR;][]{beuther16} combined with the 1.4\,\ghz\ radio
continuum data from the VLA Galactic Plane Survey
\citep[VGPS][]{stil06}.  In the identification process, we first use
the WISE Catalog of Galactic \hii\ Regions \citep{anderson14} to
separate thermal and non-thermal extended emission.  Compact sources
of radio continuum emission detected by THOR are analyzed in
\citet{bihr16} and Wang et al. (2017, in prep.).  We focus instead on
diffuse, resolved sources that are ``discrete,'' i.e. distinct from
the diffuse background emission that pervades the Galactic disk.

\section{Data}
\subsection{THOR}
THOR is a $\sim\,20$\,cm VLA survey of \hi, OH, radio recombination
line, and radio continuum emission in the Galactic plane from
$67.4\degree > \ell > 14.5\degree$, $\absb \le 1.25\degree$.  It was
conducted in VLA C-configuration, with a resolution of
$\sim\,20\arcsec$.  More survey details are given in
\citet{beuther16}. When the THOR continuum data are combined with
20\,cm VGPS continuum data, taken with the VLA in D-configuration at a
resolution of $60\arcsec$ and data taken with the 100\,m Effelsberg
telescope at a resolution of $9\arcmin$, the resulting data product is
the most sensitive radio continuum Galactic plane survey in existence
covering both large and small spatial scales.  We call this combined
data set ``THOR+VGPS.''  The THOR+VGPS data have an angular resolution
of $25\arcsec$ because of smoothing we apply to the THOR data
\citep[see][]{beuther16}.  Due to the coverage of the VGPS, the
THOR+VGPS data set is restricted to $\ell > 17.5\degree$, so our final
longitude range is $67.4\degree > \ell > 17.5\degree$.

To detect low surface brightness SNRs, the radio observations must be
sensitive to large, extended structures. To reduce confusion in the
Galactic plane, the data should also have high angular resolution.
The sensitivity of the THOR+VGPS data changes slightly over the extent
of the survey, but a typical $1\sigma$ rms value is $\sim\,1\mjyb$
\citep{bihr16}, or
$\sim 1 \times 10^{-22}$\,W\,m$^{-2}$\,Hz$^{-1}$\,sr$^{-1}$.
Over scales greater than that of the VGPS VLA D-configuration data
($\sim\,15\arcmin$), the surface brightness sensitivity should approach
that of the Effelsberg single-dish data used in the VGPS, $\sim\,1
\times 10^{-23}$\,W\,m$^{-2}$\,Hz$^{-1}$\,sr$^{-1}$ \citep{reich86,
  reich90}, although the VLA data do add some noise on large spatial
scales.
The low surface brightness noise threshold, together with the
sensitivity to small-scale structures, makes the THOR survey the ideal
data set to identify new SNRs.

\subsection{The WISE Catalog of Galactic HII Regions\label{sec:catalog}}
The WISE Catalog of Galactic \hii\ Regions \citep{anderson14} is to
date the largest, most complete catalog of \hii\ regions spanning the
entire Galaxy.  It was created by searching WISE \citep{wright10} data
by-eye for the characteristic mid-infrared (MIR) signature of
\hii\ regions: $\sim\,20$\,\micron\ emission surrounded by
$\sim\,10$\,\micron\ emission \citep[][]{anderson11}.  The
$\sim\,20$\,\micron\ emission is caused by small stochastically heated
dust grains that are mixed with the \hii\ region plasma, while the
$\sim\,10$\,\micron\ intensity is dominated by emission from
polycyclic aromatic hydrocarbons (PAHs).  All known Galactic
H{\footnotesize II} regions have this characteristic morphology.
Planetary nebulae can appear similar, but they are distinguished by
their small sizes and weak far-infrared fluxes \citep{anderson12a}.
The \hii\ region MIR emission detected by WISE and {\it Spitzer} is
about two orders of magnitude more intense than the $\sim 20$\,cm radio
continuum emission, and these observatories have sensitivities far lower
than that necessary to detect \hii\ regions across the entire Galactic
disk \citep{anderson11, anderson14}.  This single MIR morphological criterion can
therefore be used to identify all Galactic \hii\ regions.

The WISE catalog contains $\sim\,8000$ objects with the MIR morphology
of \hii\ regions, of which $\sim\,2000$ are \hii\ regions with
measured ionized gas velocities (H$\alpha$ or radio recombination
line, RRL).  This includes all known \hii\ regions, indicating that
the MIR morphological criterion can be used to identify all known
Galactic \hii\ regions.  The remaining $\sim\,6000$ sources that lack
ionized gas spectroscopic detections are \hii\ region candidates, and
there are two sub-classes: $\sim\,2000$ ``radio-loud'' candidates that
have spatially coincident radio continuum emission and $\sim\,4000$
``radio-quiet'' candidates that do not.  Radio continuum emission,
caused by the free-free emission of the ionized gas, makes the
identification of \hii\ regions more secure \citep[e.g.,][]{haslam87}.
The distribution of known regions in the catalog is statistically
complete for all \hii\ regions with ionizing fluxes consistent with
single O-stars of all spectral sub-types (Armentrout et al., 2017, in
prep., Mascoop et al., 2017, in prep.).

\subsection{Green Catalog}
G14 is the most up-to-date and authoritative catalog of Galactic SNRs.
It currently contains 294 regions compiled from the literature, and
tabulates their spatial coordinates, their 1\,\ghz\ flux densities,
spectral indices, and angular sizes.  The catalog sources cover the
entire sky, but since it is not derived from a homogeneous survey, the
catalog sensitivity varies with Galactic location.  \citet{green04}
suggest that an earlier version of the catalog than that used here was
complete to a radio surface density limit of
$10^{-20}$\,W\,m$^{-2}$\,Hz$^{-1}$\,sr$^{-1}$.  In addition to the
surface brightness limit, the catalog appears to be lacking the small
angular size SNRs that are expected \citep{green15}.
  
\section{Methodology}
Our method relies on identifying discrete regions of radio continuum
emission that a)~are not associated with \hii\ regions from the WISE
catalog and b)~lack {\it Spitzer} or WISE MIR emission.  These
criteria are somewhat redundant, as nearly all discrete sources of
coincident MIR and radio continuum emission in the Galactic plane are
\hii\ regions and are included in the WISE catalog.  We do not have a
preferred morphology for the regions we identify aside for avoiding
long filamentary radio continuum features that, based on the
morphologies of known SNRs, are not likely to be SNRs.

\begin{figure*}
  \begin{centering}
    \hskip -0.78in%
    \includegraphics[width=7.in]{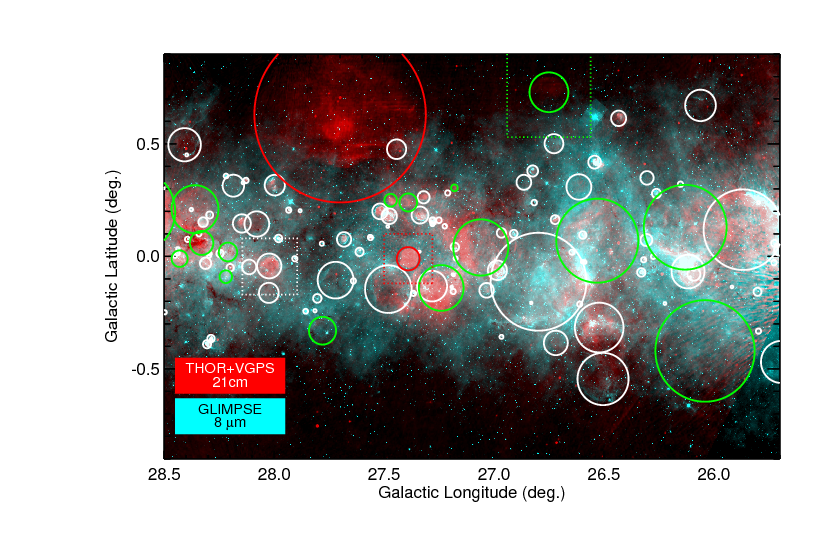}
    \vskip -15pt
    \includegraphics[width=1.88in]{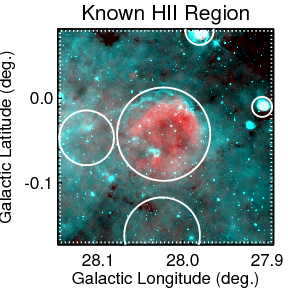}
    \includegraphics[width=1.88in]{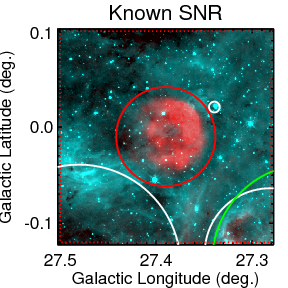}
    \includegraphics[width=1.88in]{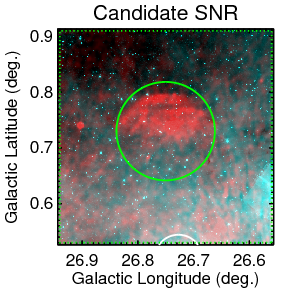}
     \caption{(Top) Two-color images with GLIMPSE 8.0\,\micron\ data
       in cyan and THOR+VGPS 21\,cm continuum data in red.
       \hii\ regions have 8.0\,\micron\ emission surrounding the
       21\,\cm emission.  Although not shown, MIPSGAL
       24\,\micron\ emission has a similar morphology as the radio
       continuum for \hii\ regions, and is essentially absent for
       SNRs.  The candidate SNRs are enclosed by green circles, known
       SNRs by red circles, and known or candidate \hii\ regions by
       white circles.   Dotted boxes enclose the areas displayed
         as insets below.  These insets show a known \hii\ region that
         has bright 8.0\,\micron\ emission (left; G028.022$-$00.043),
         a known SNR (middle; G27.4+0.0), and an example SNR candidate
         (right; G26.75+0.73).  The \hii\ region has strong
         8.0\,\micron\ emission surrounding the radio continuum, but
         the known and candidate SNRs are devoid of
         8.0\,\micron\ emission.}
     \label{fig:snrs}
  \end{centering}
\end{figure*}

To locate new SNR candidates, we search the THOR+VGPS data by-eye.  We
first identify all discrete, extended radio continuum sources that are
not associated with an \hii\ region in the WISE catalog.  This initial
search allows us to separate SNR candidates from the much more
numerous population of \hii\ regions.  We then search {\it Spitzer}
GLIMPSE 8.0\,\micron\ \citep{benjamin03, churchwell09} and MIPSGAL
24\,\micron\ \citep{carey09} data at the location of each identified
source to ensure that there is no detectable MIR emission.  These MIR
surveys have sensitivities sufficient to detect all \hii\ regions
across the entire Galaxy.  For the few sources with latitudes outside
the range of the {\it Spitzer} surveys, we use WISE 12 and
22\,\micron\ data \citep{wright10}.  Our process should remove
planetary nebulae and any remaining \hii\ regions not included in the
WISE catalog.  The remaining radio continuum sources are either SNR
candidates or known SNRs.  By matching the positions and sizes with
the G14 catalog, we determine which of these sources have been
previously identified as SNRs.  We illustrate the identification
process in Fig.~\ref{fig:snrs}.

For each identified SNR candidate, as well as all previously-known
SNRs, we compute the 1.4\,\ghz\ THOR+VGPS flux density using aperture
photometry, following the methodology of \citet{anderson12a}.  We
define a circular aperture for each source that completely contains
its radio continuum emission.  For SNR candidates that have
partial-shell morphologies, the circular aperture follows the
curvature of the visible portion of the shell.  We define four
background apertures for each source.  The background apertures sample
the local background and avoid discrete continuum sources not
associated with the SNR.  We attempt to make the background apertures
as large as possible, and to space them evenly around the source.  If
there are large-scale gradients in the background level, however, we
sample these gradients.  In complicated fields, we must define smaller
background apertures, but we still aim to space them evenly around the
source.  Five SNR candidates are low surface brightness and confused
with nearby regions, and we do not compute their flux densities.

We then compute the source integrated intensity as
\begin{equation}
  I = \frac{1}{4} \sum_{i=1}^4 I_i = \frac{1}{4} \sum_{i=1}^4 \left(I_{0} - \frac{B_i}{N_{B,i}} \times N_{S} \right)\,,
\end{equation}
and the source integrated intensity uncertainty as
\begin{equation}
\sigma_I = \sqrt{\frac{1}{4} \sum_{i=1}^4\left(I_i - I\right)^2}\,,
\end{equation}
where the summations are carried out over the four background
apertures, $I$ is the average integrated source intensity, $I_i$ is the
integrated source intensity found using one background aperture,
$I_{0}$ is integrated source intensity before background subtraction,
$B_i$ is the integrated intensity from one background aperture,
$N_{B,i}$ is the number of pixels within one background aperture, and
$N_{S}$ is the number of pixels within the source aperture.  This
method subtracts the mean intensity of a background aperture from
every pixel in the source aperture.  The derived uncertainties ignore
the approximately 20\% uncertainty in the absolute intensity
calibration of the THOR+VGPS data.

We convert $I$, which has units of \jyb, to flux densities in Jy using
the THOR+VGPS circular synthesized beam size of $25\arcsec$.  If there
are any unrelated continuum sources that fall within the source
aperture (typically extragalactic point sources or \hii\ regions), we
manually remove their flux densities from the source flux density.  We
use only the flux density values, rather than intensities, in
subsequent analyses.

There are a couple complications with our method.  First, there are
numerous filamentary features in the Galactic plane observed in radio
continuum emission.  These features are frequently located near large
massive star formation complexes.  We interpret them as being
dense thermally emitting ionized gas interacting with atomic or
molecular material in the ISM, and do not catalog such regions as
possible SNRs.  Another unrelated complication also arises around
massive star formation complexes, where bright continuum emission
produces interferometric artifacts that do not have MIR counterparts,
and therefore can be mistaken for SNRs \citep[see][their Figs.~7 and
  8]{beuther16}.  To reduce the chance of identifying artifacts, we
verify that all identified SNR candidates near large star formation
complexes are also detected in the NVSS \citep{condon98} or MAGPIS
surveys.  Due to the higher probability that a radio continuum feature
is thermally emitting ionized gas or an interferometric artifact, we
are conservative in our identifications around large star formation
complexes.

We classify the radio continuum morphology of each SNR candidate as
``shell,'' for those with well-defined radio continuum shells,
``filled,'' for those lacking an outer shell but emission filling a
roughly circular region, or ``composite'' for those that have a shell
with a filled interior.  For seven of the smallest regions, the THOR+VGPS
resolution is insufficient to determine their morphological
classification.



\section{Results}
We identify 76 new Galactic SNR candidates, and detect the radio
continuum emission from 52 of 53 previously-known SNRs from the G14
catalog.  In our aperture photometry measurements, we create a
circular aperture that encloses the radio continuum emission of each
source and therefore define the centroid and radius of each region.
We give parameters of the new SNR candidates in Table~\ref{tab:new},
which lists the Galactic longitude, Galactic latitude, and radius, as
defined in THOR+VGPS data, the 1.4\,\ghz\ THOR+VGPS flux density and
its uncertainty, the radio continuum morphological type, and the name
from H06 if the same source was identified there.  Seven SNR
candidates are so confused with nearby radio continuum sources that
their flux densities are unreliable; we do not list flux densities for
these seven regions.  We give the parameters of the G14 regions in
Table~\ref{tab:green}, which has the same columns as
Table~\ref{tab:new} but additionally contains the $1\,\ghz$ flux
density and spectral index ($\alpha$) from the G14 catalog.  We show
THOR+VGPS and MIR two-color images for example SNR candidates in
Fig.~\ref{fig:examples}, and for all candidates in the Appendix.  We
plot the Galactic locations of all known and candidate SNRs in
Fig.~\ref{fig:galdist}.

\begin{figure*}
  \begin{centering}
    \includegraphics[width=3.25in]{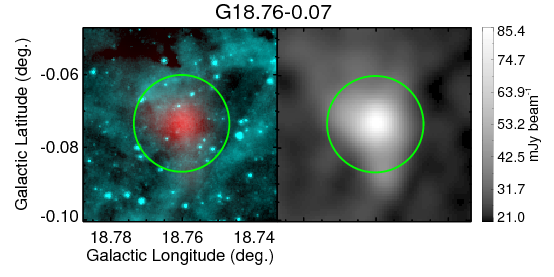}
    \includegraphics[width=3.25in]{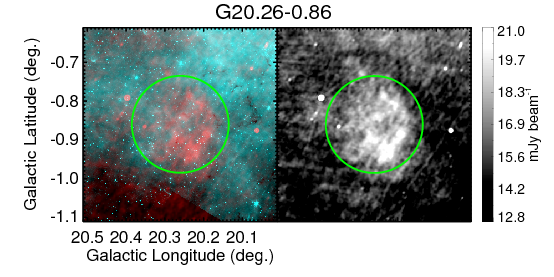}
    \includegraphics[width=3.25in]{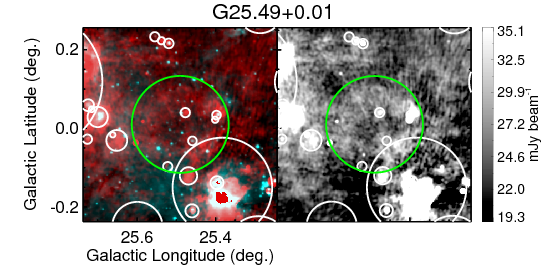}
    \includegraphics[width=3.25in]{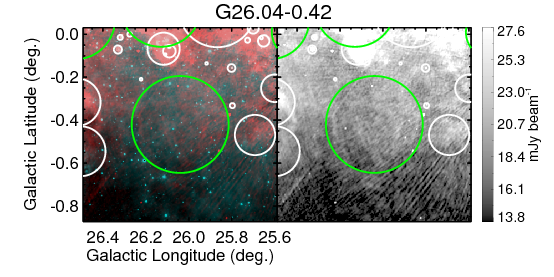}
    \includegraphics[width=3.25in]{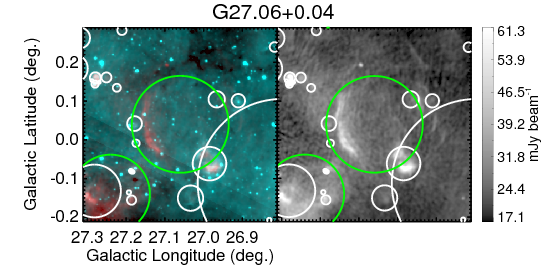}
    \includegraphics[width=3.25in]{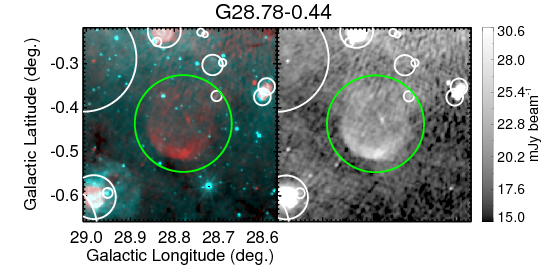}
    \caption{Example images for select SNR candidates, showing the
      range of surface brightnesses and morphologies.  The left panels
      show two-color GLIMPSE 8.0\,\micron\ (cyan) and THOR+VGPS 21\,cm
      (red), as in Fig.~\ref{fig:snrs}.  The right panels show
      THOR+VGPS data alone.  Circles in both panels are the same as in
      Fig.~\ref{fig:snrs}, with candidate SNRs are enclosed by green
      circles, known SNRs by red circles, and known or candidate
      \hii\ regions by white circles.
     \label{fig:examples}}
  \end{centering}
\end{figure*}

    \begin{figure*}
    \includegraphics[width=7in]{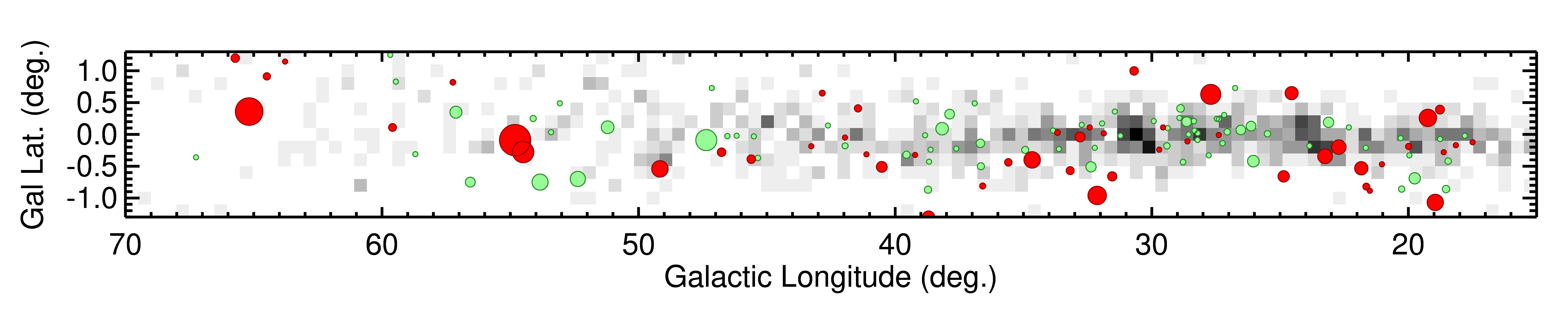}
    \caption{The Galactic distribution of the new candidate (green)
      and known (red) SNRs.  The circles approximate the SNR
      sizes, but due to the aspect ratio of the plot, the sizes are
      only valid along the Galactic longitude axis.  The background is
      a two-dimensional histogram of the \hii\ region density, with
      higher densities indicated by darker colors.\label{fig:galdist}}
    \end{figure*}

\begin{table*}
  \begin{centering}
    \scriptsize
    \setlength{\tabcolsep}{6pt}
  \caption{THOR SNR Candidates}
  \begin{tabular}{lrrrrrll}
    \hline\hline
Name & GLong & GLat & Radius$^{\rm a}$ & $S_{1.4}$ & $\sigma_{S1.4}$ & Type & H06 Name \\
 &  deg. &  deg. &  arcmin. &  Jy &  Jy & & \\
\hline
G17.80$-$0.02 & 17.800 & $-$0.020 & 4.4 & 0.29 & 0.19 & S &  \\
G18.45$-$0.42 & 18.450 & $-$0.420 & 7.6 & 2.16 & 1.77 & S &  \\
G18.53$-$0.86 & 18.530 & $-$0.860 & 8.6 & 0.43 & 0.15 & S &  \\
G18.76$-$0.07 & 18.760 & $-$0.073 & 0.8 & 0.26 & 0.04 & ? & 18.7583$-$0.0736 \\
G19.75$-$0.69 & 19.750 & $-$0.690 & 13.2 & 8.06 & 4.96 & F &  \\
G19.96$-$0.33 & 19.960 & $-$0.330 & 5.9 & 0.45 & 0.34 & C &  \\
G20.26$-$0.86 & 20.260 & $-$0.860 & 7.5 & 1.97 & 0.74 & F &  \\
G20.30$-$0.06 & 20.300 & $-$0.060 & 3.1 & 0.19 & 0.14 & S &  \\
G21.66$-$0.21 & 21.660 & $-$0.210 & 5.1 & 0.59 & 0.34 & F &  \\
G22.32+0.11 & 22.320 & 0.110 & 5.5 & 0.86 & 0.77 & S & 22.3833+0.1000  \\
G23.11+0.19 & 23.110 & 0.190 & 12.1 & \nodata & \nodata & S &  \\
G23.85$-$0.18 & 23.855 & $-$0.180 & 2.7 & 0.34 & 0.08 & S &  \\
G25.49+0.01 & 25.490 & 0.010 & 7.4 & 2.19 & 1.27 & S &  \\
G26.04$-$0.42 & 26.040 & $-$0.420 & 13.5 & \nodata & \nodata & C &  \\
G26.13+0.13 & 26.130 & 0.130 & 11.3 & 3.76 & 7.60 & S &  \\
G26.53+0.07 & 26.530 & 0.070 & 11.2 & 5.67 & 2.75 & S &  \\
G26.75+0.73 & 26.750 & 0.730 & 5.3 & 0.53 & 0.50 & F &  \\
G27.06+0.04 & 27.060 & 0.040 & 7.5 & 4.31 & 0.67 & S & 27.1333+0.0333  \\
G27.18+0.30 & 27.180 & 0.305 & 0.9 & 0.05 & 0.03 & ? &  \\
G27.24$-$0.14 & 27.240 & $-$0.140 & 6.1 & 5.36 & 1.31 & F? &  \\
G27.39+0.24 & 27.390 & 0.240 & 2.4 & 0.16 & 0.22 & F? &  \\
G27.47+0.25 & 27.467 & 0.246 & 1.7 & 0.20 & 0.11 & F? &  \\
G27.78$-$0.33 & 27.780 & $-$0.330 & 3.7 & 0.19 & 0.06 & S &  \\
G28.21+0.02 & 28.210 & 0.020 & 2.5 & 0.23 & 0.13 & F &  \\
G28.22$-$0.09 & 28.216 & $-$0.087 & 1.7 & 0.06 & 0.09 & F? &  \\
G28.33+0.06 & 28.330 & 0.060 & 3.2 & 0.42 & 0.29 & F &  \\
G28.36+0.21 & 28.360 & 0.210 & 6.4 & 2.25 & 1.93 & S & 28.3750+0.2028 \\
G28.56+0.00 & 28.564 & 0.000 & 1.5 & 0.89 & 0.10 & S & 28.5583$-$0.0083 \\
G28.64+0.20 & 28.640 & 0.200 & 11.4 & 5.90 & 4.79 & S & 28.5167+0.1333 \\
G28.78$-$0.44 & 28.780 & $-$0.436 & 6.6 & 1.63 & 1.69 & S & 28.7667$-$0.4250 \\
G28.88+0.41 & 28.880 & 0.410 & 8.9 & 1.97 & 2.26 & S &  \\
G28.92+0.26 & 28.920 & 0.260 & 3.2 & 0.34 & 0.24 & S? &  \\
G29.38+0.10 & 29.380 & 0.100 & 5.1 & 1.52 & 0.49 & C & 29.3667+0.1000 \\
G29.41$-$0.18 & 29.410 & $-$0.180 & 7.5 & 1.08 & 1.13 & S &  \\
G29.92+0.21 & 29.920 & 0.210 & 2.1 & 0.26 & 0.11 & F &  \\
G31.22$-$0.02 & 31.220 & $-$0.020 & 3.1 & 0.55 & 0.37 & S &  \\
G31.44+0.36 & 31.440 & 0.360 & 3.9 & 0.68 & 0.37 & F? &  \\
G31.93+0.16 & 31.936 & 0.172 & 2.4 & 0.23 & 0.08 & F? &  \\
G32.22$-$0.21 & 32.220 & $-$0.210 & 3.1 & 0.63 & 0.16 & F &  \\
G32.37$-$0.51 & 32.370 & $-$0.510 & 12.0 & \nodata & \nodata & S &  \\
G32.73+0.15 & 32.730 & 0.150 & 2.6 & 0.17 & 0.08 & F? &  \\
G33.62$-$0.23 & 33.620 & $-$0.230 & 2.7 & 0.26 & 0.05 & F? &  \\
G33.85+0.06 & 33.848 & 0.061 & 0.6 & 0.02 & 0.01 & ? &  \\
G34.93$-$0.24 & 34.933 & $-$0.244 & 8.1 & 0.77 & 2.37 & S &  \\
G36.66$-$0.50 & 36.660 & $-$0.500 & 8.2 & 1.29 & 1.20 & S &  \\
G36.68$-$0.14 & 36.680 & $-$0.140 & 10.0 & 2.16 & 0.58 & S &  \\
G36.90+0.49 & 36.902 & 0.488 & 3.8 & 0.50 & 0.08 & F? &  \\
G37.62$-$0.22 & 37.616 & $-$0.223 & 1.9 & 0.41 & 0.12 & F &  \\
G37.88+0.32 & 37.880 & 0.320 & 11.4 & 3.05 & 4.74 & S &  \\
G38.17+0.09 & 38.170 & 0.090 & 14.7 & \nodata & \nodata & S? &  \\
G38.62$-$0.24 & 38.620 & $-$0.240 & 2.5 & 0.10 & 0.03 & F? &  \\
G38.68$-$0.43 & 38.680 & $-$0.430 & 4.3 & 0.44 & 0.12 & F &  \\
G38.72$-$0.87 & 38.720 & $-$0.870 & 8.5 & 0.70 & 0.80 & F &  \\
G38.83$-$0.01 & 38.833 & $-$0.014 & 0.6 & 0.01 & 0.00 & ? &  \\
G39.19+0.52 & 39.190 & 0.520 & 5.5 & 0.17 & 0.21 & S? &  \\
G39.56$-$0.32 & 39.560 & $-$0.320 & 8.5 & 1.19 & 1.64 & S &  \\
G41.95$-$0.18 & 41.950 & $-$0.180 & 7.0 & 1.19 & 0.50 & S &  \\
G42.62+0.14 & 42.620 & 0.140 & 2.2 & 0.50 & 0.05 & F &  \\
G45.35$-$0.37 & 45.350 & $-$0.370 & 6.3 & 0.91 & 0.43 & F? &  \\
G45.51$-$0.03 & 45.510 & $-$0.030 & 4.1 & 1.63 & 0.42 & F? &  \\
G46.18$-$0.02 & 46.180 & $-$0.020 & 5.5 & 0.47 & 0.44 & C? &  \\
G46.54$-$0.03 & 46.540 & $-$0.026 & 6.2 & 0.85 & 0.51 & S &  \\
G47.15+0.73 & 47.150 & 0.730 & 0.8 & 0.01 & 0.00 & ? &  \\
G47.36$-$0.09 & 47.360 & $-$0.090 & 24.6 & 3.58 & 2.83 & S &  \\
G51.21+0.11 & 51.209 & 0.113 & 14.9 & 24.35 & 2.10 & ? &  \\
G52.37$-$0.70 & 52.370 & $-$0.700 & 17.7 & 5.24 & 1.75 & S &  \\
G53.07+0.49 & 53.070 & 0.490 & 1.0 & 0.06 & 0.00 & ? &  \\
G53.41+0.03 & 53.412 & 0.035 & 4.6 & 1.21 & 0.21 & S? &  \\
G53.84$-$0.75 & 53.840 & $-$0.750 & 18.7 & 1.31 & 3.43 & S? &  \\
G54.11+0.25 & 54.110 & 0.250 & 7.2 & 1.46 & 0.28 & C &  \\
G56.56$-$0.75 & 56.560 & $-$0.750 & 11.6 & 0.94 & 0.61 & F &  \\
G57.12+0.35 & 57.120 & 0.350 & 14.1 & 0.60 & 0.22 & C? &  \\
G58.70$-$0.31 & 58.700 & $-$0.310 & 4.4 & 0.16 & 0.11 & F &  \\
G59.46+0.83 & 59.460 & 0.830 & 4.5 & 0.16 & 0.03 & F &  \\
G59.68+1.25 & 59.680 & 1.250 & 5.7 & 0.25 & 0.09 & F? &  \\
G67.25$-$0.36 & 67.250 & $-$0.360 & 2.7 & 0.03 & 0.01 & F? &  \\

\label{tab:new}
\tablenotetext{a}{The radius of a circle necessary to contain
  the radio flux from the region.  For partial shells, it follows the
  curvature of the shell.}
\tablenotetext{b}{``S'' for shell-type, ``F'' for filled-center, ``C'' for composite.  Question marks (``?'') indicate uncertainty in the classification.}
  \end{tabular}
  \end{centering}
\end{table*}

\begin{table}
  \scriptsize
  \setlength{\tabcolsep}{1pt}
  \caption{G14 Known SNRs}
  \begin{tabular}{lrrrrrrrl}
\hline\hline
Name & GLong & GLat & Radius$^{\rm a}$ & $S_{1.0}^{\rm b}$ & $\alpha^{\rm b}$ & $S_{1.4}$ & $\sigma_{S1.4}$ & Type$^{\rm c}$ \\ & deg. & deg. & arcmin. & Jy &  & Jy & Jy & \\
\hline
G17.4$-$0.1 & 17.50 & $-$0.12 & 3.8 & 0.4 & $-$0.7 & \nodata & \nodata & S  \\
G18.1$-$0.1 & 18.15 & $-$0.17 & 4.6 & 4.6 & $-$0.5 & 3.82 & 0.39 & S  \\
G18.6$-$0.2 & 18.62 & $-$0.28 & 3.4 & 1.4 & $-$0.4 & 1.36 & 0.07 & S  \\
G18.8+0.3 (Kes 67) & 18.77 & 0.39 & 10.3 & 33 & $-$0.46 & 23.57 & 4.26 & S  \\
G18.9$-$1.1 & 18.95 & $-$1.07 & 19.0 & 37 & $-$0.39 & 22.75 & 7.66 & C?  \\
G19.1+0.2 & 19.24 & 0.26 & 20.2 & 10 & $-$0.5 & 12.80 & 13.06 & S  \\
G20.0$-$0.2 & 19.99 & $-$0.19 & 7.3 & 10 & $-$0.1 & 10.21 & 1.41 & F  \\
G20.4+0.1$^{\rm d}$ & \nodata & \nodata & \nodata & 9? & $-$0.1? & \nodata & \nodata & S?  \\
G21.0$-$0.4 & 21.03 & $-$0.47 & 5.6 & 1.1 & $-$0.6 & 0.84 & 0.36 & S  \\
G21.5$-$0.9 & 21.50 & $-$0.89 & 1.5 & 7 & varies & 6.36 & 0.03 & C  \\
G21.5$-$0.1$^{\rm d}$ & \nodata & \nodata & \nodata & 0.4 & $-$0.5 & \nodata & \nodata & S  \\
G21.6$-$0.8 & 21.64 & $-$0.82 & 8.1 & 1.4 & $-$0.5? & 1.18 & 0.84 & S  \\
G21.8$-$0.6 (Kes 69) & 21.83 & $-$0.53 & 15.5 & 65 & $-$0.56 & 55.96 & 7.26 & S  \\
G22.7$-$0.2 & 22.71 & $-$0.20 & 16.9 & 33 & $-$0.6 & 43.32 & 7.03 & S?  \\
G23.3$-$0.3 (W41) & 23.25 & $-$0.34 & 17.2 & 70 & $-$0.5 & 44.04 & 14.74 & S  \\
G23.6+0.3$^{\rm d}$ & \nodata & \nodata & \nodata & 8? & $-$0.3 & \nodata & \nodata & ?  \\
G24.7$-$0.6 & 24.86 & $-$0.66 & 13.1 & 8 & $-$0.5 & 4.34 & 2.27 & S?  \\
G24.7+0.6 & 24.55 & 0.65 & 15.2 & 20? & $-$0.2? & 23.48 & 7.81 & C?  \\
G27.4+0.0 (4C$-$04.71) & 27.39 & $-$0.01 & 3.1 & 6 & $-$0.68 & 3.88 & 0.24 & S  \\
G27.8+0.6 & 27.70 & 0.63 & 23.4 & 30 & varies & 33.65 & 14.71 & F  \\
G28.6$-$0.1 & 28.61 & $-$0.11 & 5.3 & 3? & ? & 5.39 & 0.58 & S  \\
G29.6+0.1 & 29.56 & 0.11 & 3.3 & 1.5? & $-$0.5? & 0.87 & 0.17 & S  \\
G29.7$-$0.3 (Kes 75) & 29.71 & $-$0.24 & 2.7 & 10 & $-$0.63 & 6.76 & 0.04 & C  \\
G30.7+1.0 & 30.69 & 1.00 & 10.0 & 6 & $-$0.4 & 3.41 & 1.69 & S?  \\
G31.5$-$0.6 & 31.54 & $-$0.66 & 10.7 & 2? & ? & 1.94 & 3.43 & S?  \\
G31.9+0.0 (3C391) & 31.87 & 0.02 & 4.5 & 25 & varies & 16.47 & 1.40 & S  \\
G32.1$-$0.9 & 32.13 & $-$0.96 & 21.5 & \nodata & ? & 1.96 & 8.46 & C?  \\
G32.4+0.1 & 32.42 & 0.11 & 4.4 & 0.25? & ? & 0.94 & 0.26 & S  \\
G32.8$-$0.1 (Kes 78) & 32.79 & $-$0.04 & 11.5 & 11? & $-$0.2? & 12.09 & 1.60 & S?  \\
G33.2$-$0.6 & 33.18 & $-$0.57 & 9.2 & 3.5 & varies & 3.47 & 1.54 & S  \\
G33.6+0.1 (Kes 79) & 33.67 & 0.03 & 6.7 & 20 & $-$0.51 & 11.05 & 1.90 & S  \\
G34.7$-$0.4 (W44) & 34.66 & $-$0.40 & 19.2 & 250 & $-$0.37 & 201.89 & 19.12 & C  \\
G35.6$-$0.4 & 35.59 & $-$0.44 & 8.6 & 9 & $-$0.5 & 9.24 & 0.17 & S?  \\
G36.6$-$0.7 & 36.59 & $-$0.81 & 7.0 & 1 & $-$0.7? & 2.10 & 1.07 & S?  \\
G38.7$-$1.3$^{\rm e}$ & 38.70 & $-$1.30 & 14.3 & \nodata & ? & \nodata & \nodata & S  \\
G39.2$-$0.3 & 39.22 & $-$0.32 & 4.5 & 18 & $-$0.34 & 11.54 & 0.68 & C  \\
G40.5$-$0.5 & 40.52 & $-$0.51 & 12.5 & 11 & $-$0.4 & 8.03 & 3.02 & S  \\
G41.1$-$0.3 (3C397) & 41.12 & $-$0.31 & 2.9 & 25 & $-$0.5 & 11.73 & 0.50 & S  \\
G41.5+0.4 & 41.45 & 0.41 & 8.5 & 1? & ? & 5.15 & 1.23 & S?  \\
G42.0$-$0.1 & 41.95 & $-$0.05 & 5.9 & 0.5? & ? & 0.96 & 0.36 & S?  \\
G42.8+0.6 & 42.84 & 0.65 & 6.9 & 3? & $-$0.5? & \nodata & \nodata & S  \\
G43.3$-$0.2 (W49B) & 43.27 & $-$0.19 & 3.2 & 38 & $-$0.46? & 26.64 & 0.65 & S  \\
G45.7$-$0.4 & 45.61 & $-$0.39 & 10.0 & 4.2? & $-$0.4? & 4.28 & 1.28 & S  \\
G46.8$-$0.3 (HC30) & 46.77 & $-$0.28 & 10.0 & 17 & $-$0.54 & 14.07 & 0.83 & S  \\
G49.2$-$0.7 (W51) & 49.17 & $-$0.54 & 19.2 & 160? & $-$0.3? & 115.41 & 15.09 & S?  \\
G54.1+0.3$^{\rm d}$ & \nodata & \nodata & \nodata & 0.5 & $-$0.1 & \nodata & \nodata & C?  \\
G54.4$-$0.3 (HC40) & 54.50 & $-$0.28 & 25.0 & 28 & $-$0.5 & 21.06 & 5.99 & S  \\
G55.0+0.3 & 54.81 & $-$0.09 & 36.5 & 0.5? & $-$0.5? & 14.20 & 6.13 & S  \\
G57.2+0.8 (4C21.53) & 57.24 & 0.82 & 6.7 & 1.8 & $-$0.62 & 1.34 & 0.07 & S?  \\
G59.5+0.1 & 59.59 & 0.11 & 9.0 & 3? & ? & 1.79 & 0.84 & S  \\
G59.8+1.2$^{\rm d}$ & \nodata & \nodata & \nodata & 1.5 & 0 & \nodata & \nodata & ?  \\
G63.7+1.1 & 63.78 & 1.15 & 4.8 & 1.8 & $-$0.24 & 1.60 & 0.03 & F  \\
G64.5+0.9 & 64.49 & 0.91 & 8.4 & 0.15? & $-$0.5 & 0.24 & 0.07 & S?  \\
G65.1+0.6 & 65.18 & 0.36 & 32.1 & 5.5 & $-$0.61 & 6.15 & 2.19 & S  \\
G65.7+1.2 (DA495) & 65.72 & 1.20 & 9.9 & 5.1 & varies & 2.86 & 0.30 & F  \\
G65.8$-$0.5$^{\rm d}$ & \nodata & \nodata & \nodata & \nodata & ? & \nodata & \nodata & S  \\
G66.0+0.0$^{\rm e}$ & \nodata & \nodata & \nodata & \nodata & ? & \nodata & \nodata & S  \\

\label{tab:green}
\tablenotetext{a}{The radius of a circle necessary to contain the
  radio flux from the region.  For partial shells, it follows the
  curvature of the shell.}
\tablenotetext{b}{Question marks (``?'') indicate that the value in the G14 catalog is uncertain.}
\tablenotetext{c}{``S'' for shell-type, ``F'' for filled-center, ``C'' for composite.  Question marks (``?'') indicate uncertainty in the classification.}
\tablenotetext{d}{HII region (see text).}
\tablenotetext{e}{Not detected in THOR+VGPS continuum data.}
  \end{tabular}
  \end{table}

\subsection{New SNR Candidates}
All 76 SNR candidates have THOR+VGPS 21\,cm continuum emission and a
deficiency of MIR emission compared with \hii\ regions.
Of the 76 candidates, seven were identified previously as being
possible SNRs in H06, and one was identified in B06.  These works
utilized the same MIR deficit as we use here, but also employed data
from multiple radio frequencies in an effort to determine the spectral
indices.  Our identifications therefore provide some additional support
to the object being a true SNR, although this support is limited due
to the similarities between our methodologies.

THOR includes multiple continuum spectral windows that in principle
allow for the computation of spectral indices.  Because the VGPS data
only have one continuum spectral window, however, we cannot compute
spectral indices for SNR candidates that require THOR+VGPS data to be
detected.  \citet{bihr16} did detect compact emission toward several
smaller SNR candidates in individual THOR spectral windows in the
first half of the THOR survey, and several more were detected in the
second half of the survey (Y.~Wang et al., 2017, in prep.).  All
candidates detected in THOR data alone have negative spectral indices
consistent with them being true SNRs.

\subsection{G14 SNRs}
We confirm the radio emission for 52 G14 SNRs that lie within the
THOR+VGPS zone, but we did not detect THOR+VGPS radio emission from
G66.0$-$0.0.  This region was detected in radio continuum emission by
\citet{sabin13} in the GB6 5\,\ghz\ data \citep{gregory96}.  They note
that it was not detected at 20\,cm, which we confirm in the THOR+VGPS
data.  The nature of this source is therefore unclear.

\begin{figure}
  \begin{centering}
  \includegraphics[width=3in]{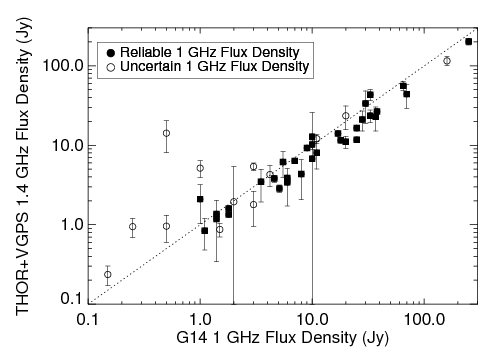}
  \caption{THOR+VGPS 1.4\,\ghz\ flux densities compared with G14
    1\,\ghz\ flux densities.  Filled circles denote SNRs that have more
    secure values in the G14 catalog, whereas open circles are more
    uncertain (values that have a question mark in the G14 catalog).
    The dotted line shows a 1:1 relationship.\label{fig:fluxes}}
    \end{centering}
\end{figure}

The flux densities derived from our aperture photometry measurements
agree well with those listed in the G14 catalog, as illustrated by
Fig.~\ref{fig:fluxes}.  The G14 catalog contains flux densities at
1\,\ghz, extrapolated from the values measured using the derived
spectral index, but does not contain flux density uncertainties.
Some 1\,\ghz\ flux densities are marked as being uncertain in the G14
catalog, and these have especially large discrepancies with the
THOR+VGPS values.  That the relationship is so close to 1:1 is
evidence that the THOR+VGPS data are well-calibrated relative to
previous measurements in the literature.

\begin{figure*}
  \begin{center}
  \includegraphics[width=2.in]{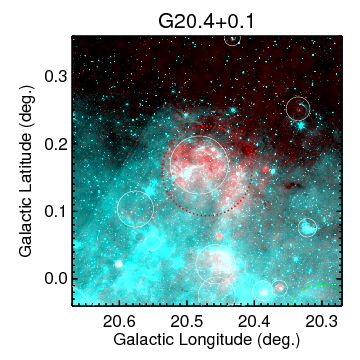}
  \includegraphics[width=2.in]{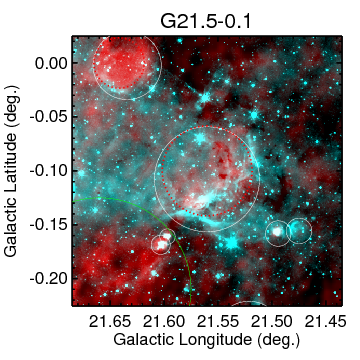}
  \includegraphics[width=2.in]{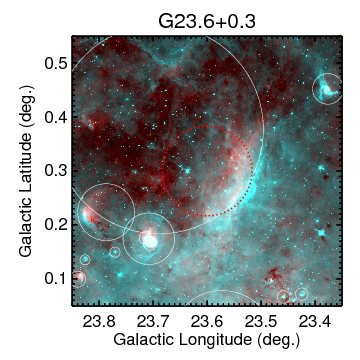}\\
  \includegraphics[width=2.in]{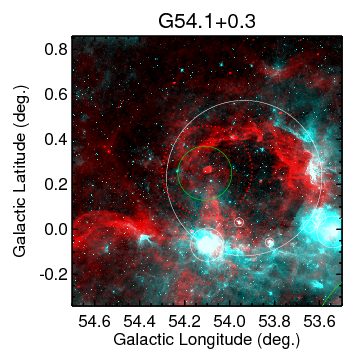}
  \includegraphics[width=2.in]{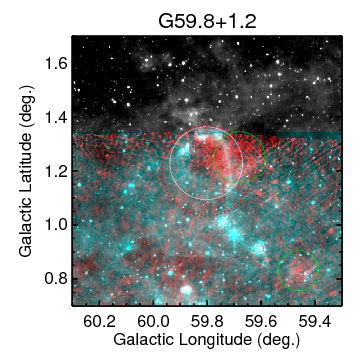}
  \includegraphics[width=2.in]{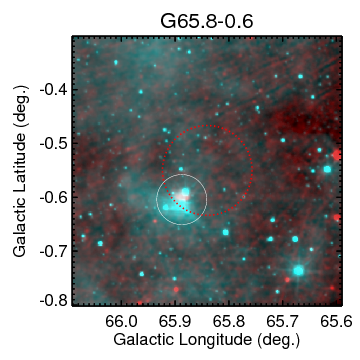}
  \caption{G14 SNRs confused with \hii\ regions.  Shown for each
    region are two-color images with GLIMPSE 8.0\,\micron\ (for
    G20.4+0.1, G21.5$-$0.1, G23.6+0.3, and G54.1+0.3) or WISE
    12\,\micron\ data (for G59.8+1.2 and G65.8$-$0.5) in cyan and
    THOR+VGPS 21\,cm continuum data in red.  The central positions are
    taken from the G14 catalog, and the image dimensions are three
    times the G14 SNR diameters.  As in Fig.~\ref{fig:snrs}, the
    candidate SNRs are enclosed by green circles and known or
    candidate \hii\ regions by white circles.  Red dotted circles
    show the expected position and size of the G14 SNRs that are
    confused with \hii\ regions.\label{fig:notreal}}
  \end{center}
\end{figure*}

Six G14 SNRs appear to be confused with \hii\ regions: G20.4+0.1,
G21.5$-$0.1, G23.6+0.3, G54.1+0.3, G59.8+1.2 and G065.8$-$0.5.  We
show THOR+VGPS data, as well as GLIMPSE 8.0\,\micron\ or WISE
12\,\micron\ data, for these regions in Fig.~\ref{fig:notreal}.  These
two infrared data sets exhibit the same morphology for \hii\ regions
\citep{anderson12a}.  We use the WISE data only if the GLIMPSE
coverage is not sufficient.  We discuss the individual regions below.

\paragraph{G20.4$+$0.1}:
The spectral index for G20.4+0.1 was found by B06 to be $-0.4$, but a
value of $-0.08\pm0.09$ was derived by \citet{sun11}.  Additionally,
\citet{pinheiro11} measured a high MIR to radio flux ratio consistent
with that of \hii\ regions.  It is spatially coincident with WISE
\hii\ region G020.482+00.167, which has measured RRL emission from
\citet{lockman89}, providing further evidence that the radio emission
is thermal.  The radio continuum emission extends to the west of the
identified \hii\ region enclosed by 8.0\,\micron\ emission.  Based on
its association with MIR emission, this western extension appears to
also be thermal, possibly resulting from photons leaking from the
\hii\ region \citep[see][]{luisi16}.

\paragraph{G21.5$-$0.1}:
The radio emission of G21.5$-$0.1 is entirely spatially coincident
with the WISE \hii\ region G021.560$-$00.108, and is bordered by
GLIMPSE 8.0\,\micron\ emission.  The \hii\ region has measured RRL
emission \citep{anderson15a}, which suggests that the radio continuum
emission is thermal. Its morphology and the high MIR flux further
point to this being an \hii\ region. Using MIPSGAL 24\,\micron\ data,
\citet{pinheiro11} also found a high MIR to radio flux ratio for this
region consistent with that of \hii\ regions.

\paragraph{G23.6$+$0.3}:
This source has radio emission along a bright linear feature.  This
feature is inside the GLIMPSE 8.0\,\micron\ emission, again indicating
that the region is an \hii\ region.  This defines a portion of the
shell of the WISE \hii\ region G23.689+00.377, which has measured RRL
emission \citep{lockman96}.  Similar to G21.5$-$0.1,
\citet{pinheiro11} found a high MIR to radio ratio for G21.5$-$0.1.
On this basis they suggested that is an \hii\ region.

\paragraph{G54.1$+$0.3}:
\citet{lang10} noted that the extended radio emission in the field of
G54.1+0.3 is ambiguous: it could be non-thermal emission associated
with the pulsar wind nebula G54.1+0.3 (the compact object at the
center of the image shown in Fig.~\ref{fig:notreal}) or could be
thermal emission.  They further note that 24\,\micron\ MIPSGAL
emission has a similar morphology as the radio loop, hinting that it
may be thermal emission.  We support the latter interpretation, and
associate the radio continuum emission with the WISE \hii\ region
G053.935+00.228.  On its western edge there is strong
8.0\,\micron\ GLIMPSE emission, and the radio emission is found
interior to this MIR emission.  \citet{lockman89} measured RRL
emission from a position on the western edge.  Together, these data
indicate that the radio emission previously suggested as being a
possible SNR remnant associated with the G54.1+0.3 pulsar wind nebula
(PWN) actually represents a thermal \hii\ region.  Importantly,
however, there is diffuse, extended radio emission that has a
different morphology from the ring structure.  This emission is faint,
centered approximately on the compact PWN, and has a radius of
$5\arcmin$.  We include it as a candidate SNR.

\paragraph{G59.8$+$1.2}:
Although G59.8+1.2 is almost certainly an \hii\ region, there is a
nearby patch of radio emission that, based on its lack of MIR
emission, does appear to be nonthermal.  We identify this non-thermal
emission as SNR candidate G59.68+1.25.  These two regions are perhaps
confused in the literature.  We suggest that the G59.8+1.2 object is
actually the WISE \hii\ region G059.803+01.228, which has measured RRL
emission (L.~Anderson et al., 2017, in prep.). The region was also
listed as having a flat spectral index by \citet{sun11}, who
additionally note that more observations are required to establish its
classification, in support of it being confused with the nearby
\hii\ regions.

\paragraph{G65.8$-$0.5}:
The case of G65.8$-$0.5 is especially confusing.  Much of the
H-$\alpha$ emission mentioned in \citet{sabin13} is associated with
the compact WISE \hii\ region candidate G065.887$-$00.605.  This region
is a candidate because although it has the characteristic MIR and
radio continuum morphology of \hii\ regions, it has not been measured
in RRL or H-$\alpha$ spectroscopic observations.  \citet{sabin13} show
low-frequency radio data with a slightly extended morphology not
present in the THOR+VGPS data.  Given that the low-frequency radio
data peak intensity is spatially coincident with the \hii\ region, it
seems likely that all the radio emission is thermal.

\subsection{H06 and B06 Mis-identifications}
\begin{figure*}
    \begin{center}
\includegraphics[width=1.75in]{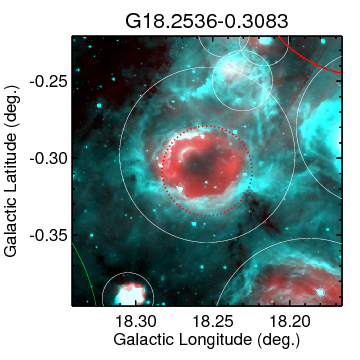}
\includegraphics[width=1.75in]{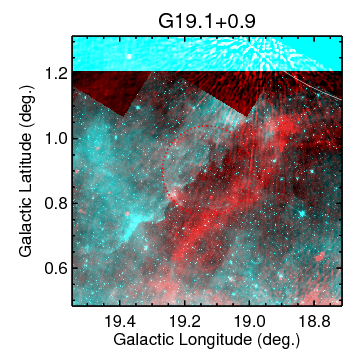}
\includegraphics[width=1.75in]{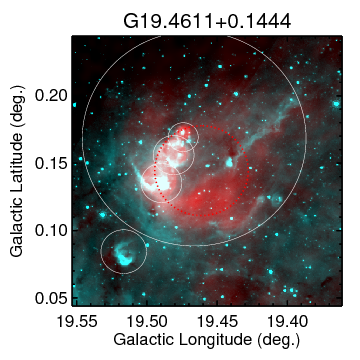}
\includegraphics[width=1.75in]{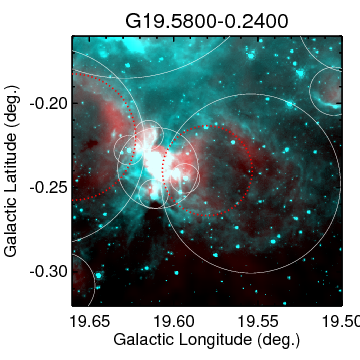}
\includegraphics[width=1.78in]{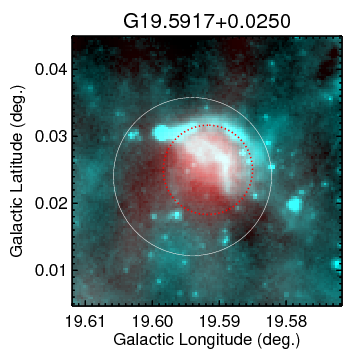}
\includegraphics[width=1.78in]{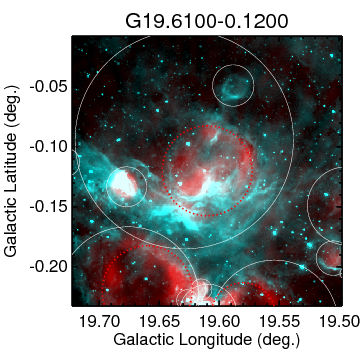}
\includegraphics[width=1.78in]{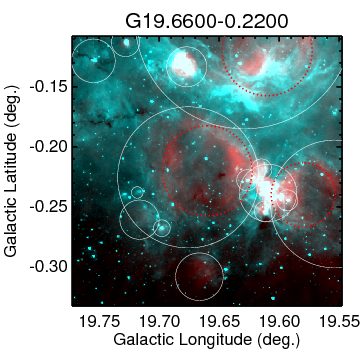}
\includegraphics[width=1.78in]{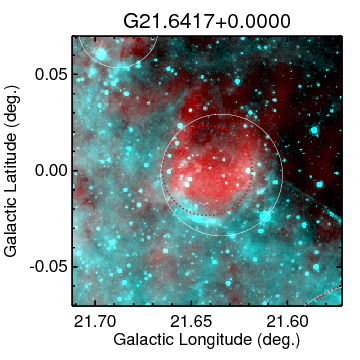}
\includegraphics[width=1.78in]{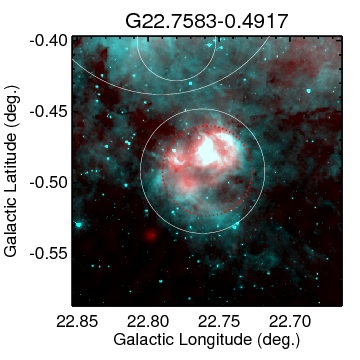}
\includegraphics[width=1.78in]{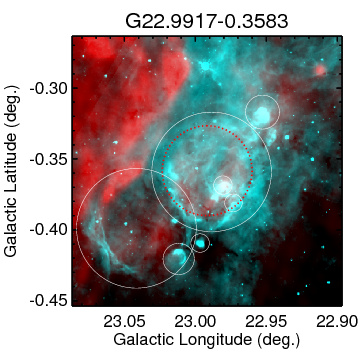}
\includegraphics[width=1.78in]{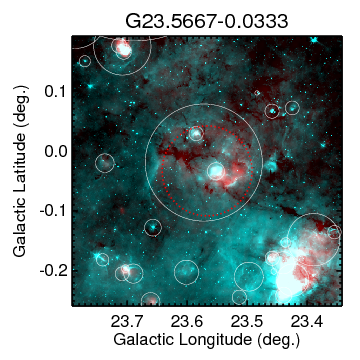}
\includegraphics[width=1.78in]{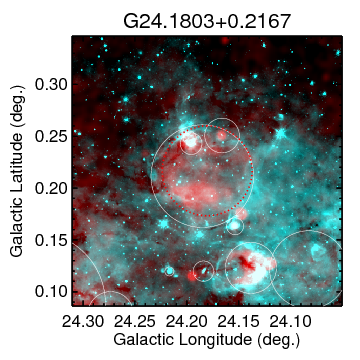}
\includegraphics[width=1.78in]{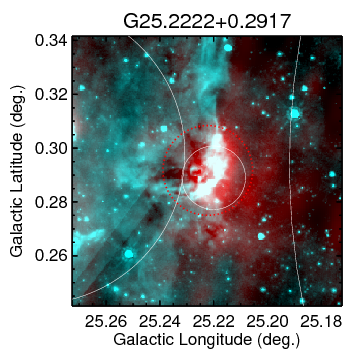}
\includegraphics[width=1.78in]{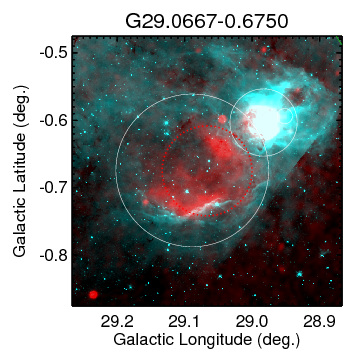}
\includegraphics[width=1.78in]{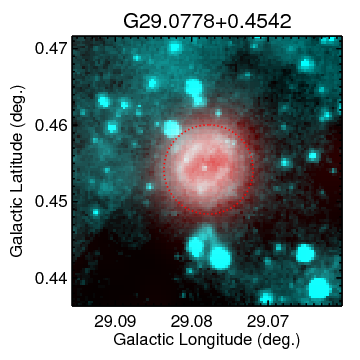}
\includegraphics[width=1.78in]{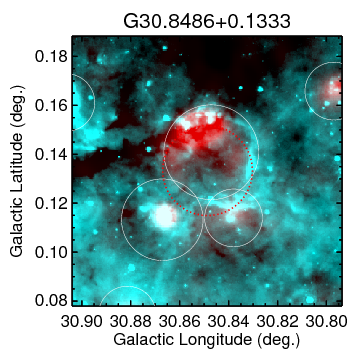}
\includegraphics[width=1.78in]{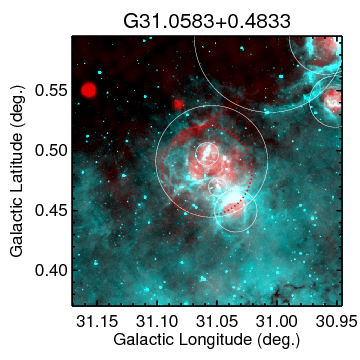}
\includegraphics[width=1.78in]{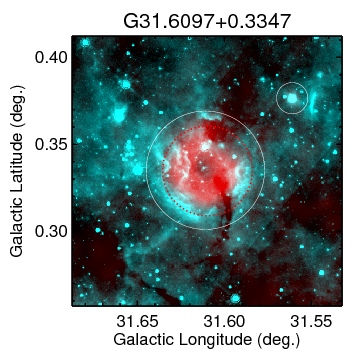}
\includegraphics[width=1.78in]{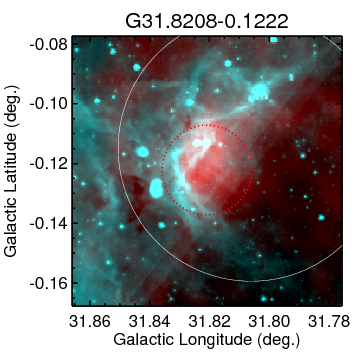}
\caption{Mis-identified SNR candidates from H06 and B06.  The format
  for these images is the same as that of Fig.~\ref{fig:snrs}, with
  GLIMPSE 8.0\,\micron\ data in cyan and THOR+VGPS 21\,cm continuum
  data in red.  As in Fig.~\ref{fig:snrs}, the candidate SNRs are
  enclosed by green circles, known SNRs by red circles, and known or
  candidate \hii\ regions by white circles.  Red dotted circles show
  the expected position and size of the SNR candidates.  All except
  for G19.1+0.9, which appears to be a radio continuum filament,
  and G29.0778+0.4542, which is a planetary nebula, are confused with
  \hii\ regions. \label{fig:bad2}}
\end{center}
\end{figure*}

Many SNR candidates identified in the MAGPIS survey of H06 are
actually \hii\ regions.  Additionally, we suggest that one B06 SNR
candidate (G19.13+0.90) is a thermally-emitting filament.  We show the
mis-identified MAGPIS regions, as well as G19.13+0.90, in
Fig.~\ref{fig:bad2}.

The 17 mis-identified MAGPIS SNR candidates are: G18.2536$-$0.3083,
G19.4611+0.1444, G19.5800$-$0.2400, G19.5917+0.0250,
G19.6100$-$0.1200, G19.6600$-$0.2200, G21.6417+0.0000,
G22.7583$-$0.49171, G22.9917$-$0.3583, G23.5667$-$0.0333,
G24.1803+0.2167, G25.2222+0.2917, G29.0667$-$0.6750, G30.8486+0.1333,
G31.0583+0.4833, G31.6097+0.3347, and G31.8208$-$0.1222.  All are
spatially coincident with a known \hii\ region from the WISE catalog.
Of these, G18.254-0.308 was previously mentioned in \citet{bihr16} as
being an \hii\ region.  One additional MAGPIS SNR (G29.0778+0.4542) is
a known planetary nebula (PN~A66~48).

The source G19.13+0.90 from B06 does not appear to be a true SNR.  B06
classify this object as ``class III,'' their lowest certainty of
actually being a SNR.  It has associated MIR emission, and its radio
morphology is that of a long filament.  Although it does have a
spectral index of $-0.5$ reported in B06, the morphology and MIR
emission from this feature makes its classification as a SNR
uncertain.

\subsection{Comparison with known SNRs and HII regions}
\begin{figure*}
  \begin{centering}
    \includegraphics[width=3in]{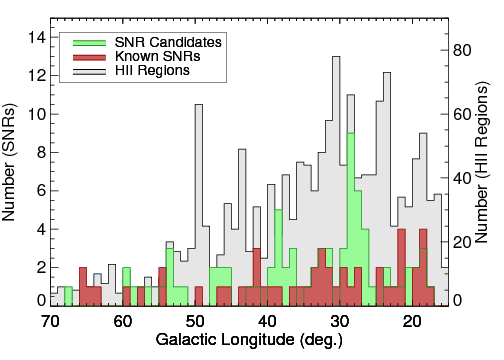}
    \includegraphics[width=3in]{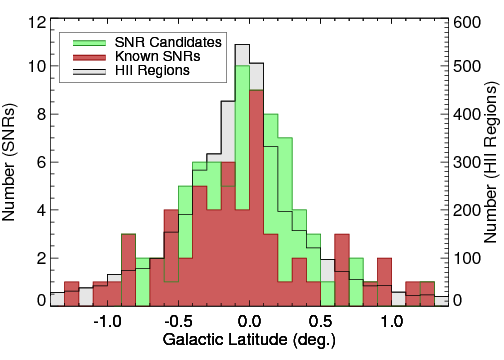}
    \caption{Galactic longitude (left) and latitude size (right)
      distributions for candidate (green) and known (red) SNRs. The
      light gray histograms show the distribution of \hii\ regions
      from the WISE catalog.  The longitude distributions of the
      candidate and known SNRs are similar, with the most obvious
      exception being the larger number of candidates near $\ell =
      30\degree$.  The Galactic latitude distributions in the right
      panel are not statistically different, nor are they different
      from that of \hii\ regions.  There is no indication that the SNR
      sample is significantly incomplete near $b = 0\degree$ where
      confusion is the greatest.\label{fig:glongglat}}
  \end{centering}
\end{figure*}

The Galactic longitude distribution of the SNR candidates shown in the
left panel of Fig.~\ref{fig:glongglat} is similar to that of the
previously known sample.  There are, however, larger numbers of
candidate SNRs near $\ell \simeq 30\degree$ compared with the G14
population.  The long bar ends at $\ell \simeq 30\degree$
\citep{benjamin05}, and the giant \hii\ region W43 is at this
longitude.  There is also a greater number of star formation regions
at $\ell \simeq 30\degree$, as seen in the WISE \hii\ region
distribution in Figs.~\ref{fig:galdist}~and~\ref{fig:glongglat}.  The
increased number of SNRs here is consistent with the large star
formation rate of the W43 region.
Kolmogorov-Smirnov (K-S) tests show that the candidate SNR, G14
SNR, and \hii\ region populations are consistent with originating from
the same parent distribution.  We use a K-S probability threshold of
0.01 for this and all subsequent calculations.
  
The discrepancy between the number of SNRs and the number of
\hii\ regions may give some clues about the recent star formation
rate.  While \hii\ regions primarily trace younger O-stars, SNRs arise
from both O and B-stars.  Since B-stars are on average older than
O-stars, SNRs trace an older stellar population than \hii\ regions.
Young massive star formation regions are thus more likely to have a
higher ratio of \hii\ regions to SNRs.  For example, the W49 region
near $\ell = 43\degree$ is known to have an age of $\sim\,1.5$\,Myr
\citep{wu16}, which is less than the lifetime of the most massive
O-stars.  The large number of \hii\ regions here relative to the SNR
population is consistent with this interpretation.  There are similar
discrepancies near $\ell = 49\degree$, associated with W51, and $\ell
= 25\degree$.  It is worth noting that W51 does host a large, luminous
SNR G49.2$-$0.7, implying that there were multiple generations of star
formation in the region. In contrast, there is no single large
\hii\ region complex near $l=25\degree$.

The Galactic latitude distributions in the right panel of
Fig.~\ref{fig:glongglat} are similar for the known and candidate SNRs.
Both distributions are peaked near $b = 0\degree$, but the candidates
are skewed more toward positive latitudes compared with the G14
regions.  K-S tests show that these differences are not significant,
and the latitude distributions of the two populations are not
statistically different from that of \hii\ regions from the WISE
catalog.  Based on the similarity of the \hii\ region and SNR latitude
distributions, there is no indication that the SNR distribution is
missing a significant number of sources near $b=0\degree$ where
confusion is greatest.

Our distribution of morphological types is heavily skewed toward
filled-center morphologies compared with the G14 sample.  Of the 52
G14 SNRs in the survey zone, over 75\% (41 of 52) are classified as
having a shell morphology (including uncertain designations of ``S?'';
see Table~\ref{tab:green}).  About 15\% are classified as having
composite morphologies (7 of 52) and $\sim 10\%$ are classified as
having filled-center morphologies (4 of 52).  About half of the SNR
candidates with morphological classifications (34 of 69) have shell
morphologies, whereas about 40\% have filled center morphologies (29 of
69) and about 10\% have composite morphologies (6 of 69).  The
comparatively high percentage of filled-shell SNR
candidates may be due to our detection method, which is less biased
against a particular morphological type.

\begin{figure}
  \includegraphics[width=3in]{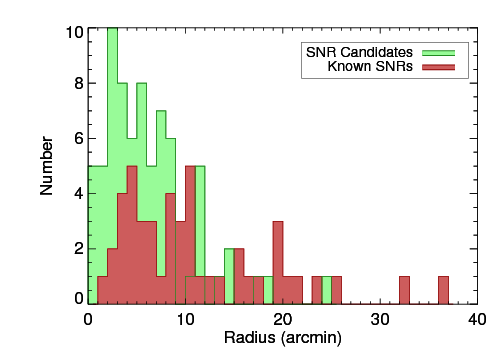}   
  \caption{Angular radius distributions for candidate (green) and
    known (red) SNRs.  The candidate SNRs are on average smaller than
    the known SNRs.  \label{fig:size}}
\end{figure}

As shown in Fig.~\ref{fig:size}, the candidate SNR angular radius
distribution skews toward lower values compared with the distribution
for known SNRs.  This is expected given the higher resolution
THOR+VGPS data compared with previous radio surveys.  The average and
standard deviation of the candidate and known SNR radii are
$6.4\arcmin\pm 4.7\arcmin$ and $11.0\arcmin\pm 7.8\arcmin$,
respectively.  The two radius distributions are however consistent
with originating from a single parent distribution, according to a K-S
test.  Many of the larger known and candidate SNRs are at higher
Galactic longitudes (Fig.~\ref{fig:galdist}).  Although this may be
due to selection effects arising from confusion at lower longitudes,
it may also be explained by the fact that the mean heliocentric
distance is smaller at higher longitudes.

The angular sizes of the candidate SNRs are on average larger at
higher Galactic longitudes.  For example the average radius for SNR
candidates with $\ell > 35\degree$ is $7.6\arcmin$, whereas it is
$5.6\arcmin$ for candidates with $\ell < 35\degree$.  The effect may
be real, or may be caused by selection effects.  For example, this may
indicate that we are unable to identify larger regions in more
confused regions at lower Galactic longitudes, or that the angular
sizes increase at higher longitudes because the mean heliocentric
distance decreases.


\begin{figure}
  \begin{centering}
  \includegraphics[width=3in]{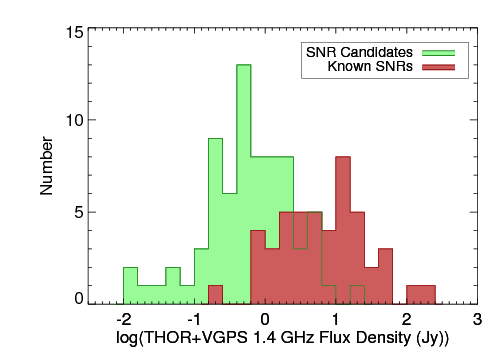}
  \caption{The flux density distribution of new candidate (green)
    and known (red) SNRs, as measured in the THOR+VGPS
    data.  The distribution of flux densities for the new
    candidates is shifted toward significantly lower values compared
    with that of the known sample.\label{fig:fluxdist}}
  \end{centering}
  \end{figure}

The distribution of flux densities derived from THOR+VGPS data for the
new SNR candidates is shifted toward significantly lower values
compared with the SNRs in the G14 catalog (Fig.~\ref{fig:fluxdist}).
The average and standard deviation of the base-ten logarithm of the
G14 sources is $0.82\pm0.61$, while it is $-0.23\pm0.65$ for the
candidates.  A K-S test shows that these differences are statistically
significant.  The lowest flux G14 sources have flux densities near
1\,\jy, although 71\% of the candidates have flux densities less than
1\,\jy.  This result is unsurprising, given that a main advantage of
the THOR+VGPS data is that they are more sensitive than previous data.
Future surveys, for example with MeerKat, ASKAP, MWA, and LOFAR, will
undoubtedly discover additional regions.

\begin{figure}
  \begin{center}
  \includegraphics[width=3.in]{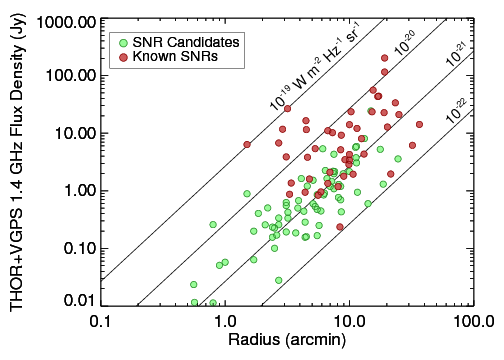}
  \caption{Flux density versus radius for G14 (filled circles) and
    candidate (plus signs) SNRs.  The solid lines show surface
    brightness limits from $10^{-19}$ to $10^{-22}$\,
    \,W\,m$^{-2}$\,Hz$^{-1}$\,sr$^{-1}$.  The surface brightness
    therefore decreases from the top left of the plot to the bottom
    right.  The surface brightness distribution of the THOR SNR
    candidates is similar to that of the known
    SNRs.\label{fig:radius_flux}}
  \end{center}
\end{figure}

To investigate whether the detection of the new SNR candidates is due
to higher sensitivity or better angular resolution, we examine the
flux density versus radius for the G14 and candidate SNRs in
Fig.~\ref{fig:radius_flux}.  This figure shows that the two
populations separate primarily by intensity (see also
Fig.~\ref{fig:fluxdist}), but also by radius (see also
Fig.~\ref{fig:size}).  Furthermore, all the low-intensity, small
radius data points are from THOR SNR candidates.

Surface brightness decreases in Fig.~\ref{fig:radius_flux} toward the
bottom right.  There are fractionally more SNR candidates that have
lower average surface brightness values compared with the known SNRs.
For example, almost half of the candidates fall between the $10^{-21}$
and $10^{-22}$\,W\,m$^{-2}$\,Hz$^{-1}$\,sr$^{-1}$ lines whereas only
$\sim\,10\%$ of the known SNRs do.  There are nevertheless many G14
SNRs with low surface brightness values similar to those of the THOR
candidates.  The main advantage of THOR over previous radio continuum
data for the present inner-Galaxy study is its improved angular
resolution, which allows us to better identify small sources and
the thin shells of large sources in complicated fields.

\subsection{Implications for the Galactic SNR population}
If confirmed, the 76 new SNR candidates would more than double the
known SNR population in the survey zone (52 to 128, not counting the
six objects in G14 confused with \hii\ regions), and will have
increased the total Galactic SNR population by 27\% (288 to 364, again
not counting the six regions).  Outside of the Galactic zone surveyed
in B06, we identified 67 SNR candidates, two and a half times the 45
known SNRs.  The surface brightness limit of the THOR+VGPS data is
$\sim\,10^{-22}$\,W\,m$^{-2}$\,Hz$^{-1}$\,sr$^{-1}$, which given an
expected spectral index for SNRs of $\sim\,-0.5$ is roughly equivalent
to that of B06.  The THOR+VGPS data are, however, sensitive to a wider
range of size scales than the data of B06.

The increase in SNRs suggested by our study is in rough agreement with
the prediction of B06.  Based on the assumption that future surveys
with a similar surface brightness sensitivity would also triple the
number of SNRs, B06 predicted there would be $\sim\,500$ SNRs detected
in the Galaxy.  Even this estimate of $\sim\,500$ is about a factor of
two less than expected (see Section~\ref{sec:intro}).
We agree with B06 that future, more sensitive surveys may resolve the
tension, although confusion in the inner Galaxy may limit new
detections.  Our analysis of Fig.~\ref{fig:radius_flux} suggests that
any future inner-Galaxy survey should prioritize high-angular
resolution to reduce confusion.


\section{Summary}
Using 1.4\,\ghz\ continuum data from The \hi, OH, and RRL (THOR) VLA
survey of the first quadrant of the Galactic plane, combined with
continuum data from the VLA Galactic Plane Survey (VGPS), we cataloged
76 new Galactic supernova remnants (SNRs).  Our method identifies
diffuse radio continuum emission regions lacking mid-infrared
counterparts seen for \hii\ regions and planetary nebulae.  All
candidates lack MIR emission from known \hii\ regions, as cataloged in
the WISE Catalog of Galactic \hii\ Regions.  The detected candidates
follow a similar spatial distribution compared to the previously known
sample, albeit with a larger concentration near $\ell = 30\degree$.
The low number of known and candidate SNRs near $\ell \simeq
49\degree$ (associated with W51), $43\degree$ (associated with W49),
and $25\degree$ relative to the number of \hii\ regions indicates the
relative youth of these star formation regions.  The sizes of the new
SNR candidates are on average smaller than those of the known regions
and the candidate fluxes are on average lower than those of previously
known SNRs.

We also detect radio continuum emission from 52 known SNRs from
\citet{green14a}, but fail to detect one SNRs in the surveyed region
that has a previous radio continuum detection.  We note that six known
SNRs are confused with \hii\ regions, and that a further 17 SNR
candidates from B06 and H06 are confused with \hii\ regions.  These
results show that our method is useful for classifying SNRs
mis-identified previously.

If our candidates prove to be true SNRs, they would more than double
the Galactic SNR population in the surveyed region.  Even with this
large number of new candidates, there is still a factor of two
disagreement between the number of SNRs detected and the number
expected.  Similar studies in other parts of the Galaxy using data as
sensitive as THOR could further reduce the discrepancy, although
confusion in the inner Galaxy is likely to make subsequent detections
increasingly difficult.  To maximize SNR detections, such future
surveys of the inner Galaxy should also have high angular resolution.

\acknowledgements \nraoblurb~ H.B. and Y.W. acknowledge support from
the European Research Council under the Horizon 2020 Framework Program
via the ERC Consolidator Grant CSF-648505.  R.S.K.  and
S.C.O.G. acknowledge support from the European Research Council via
the ERC Advanced Grant STARLIGHT (project number 339177). They furthermore thank the DFG for financial help via SFB 881
``The Milky Way System'' (subprojects B1, B2, and B8) and SPP 1573
``Physics of the Interstellar Medium.''  F.B. acknowledges support
from DFG grant BI 1546/1-1.

\bibliographystyle{apj}
\bibliography{ref.bib}

\begin{appendix}
\section{SNR Images}
Here we provide THOR+VGPS 1.4\,\ghz\ images for the individual
candidate SNRs.

\begin{figure*}
\includegraphics[width=3.6in]{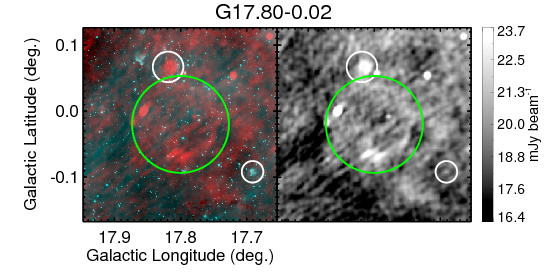}
\includegraphics[width=3.6in]{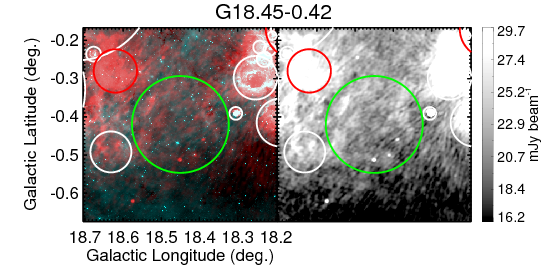}
\includegraphics[width=3.6in]{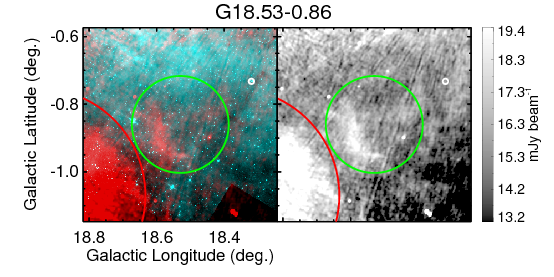}
\includegraphics[width=3.6in]{{G18.76-0.07_THOR_two}.png}
\includegraphics[width=3.6in]{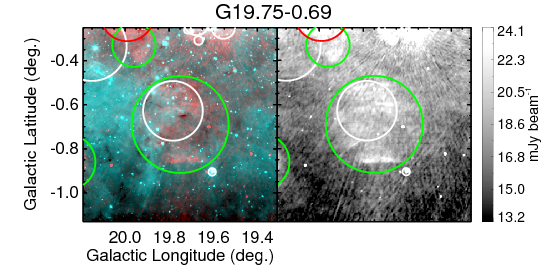}
\includegraphics[width=3.6in]{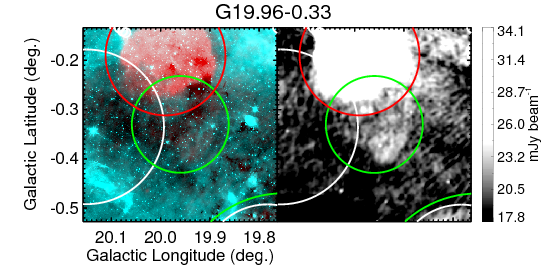}
\includegraphics[width=3.6in]{{G20.26-0.86_THOR_two}.png}
\includegraphics[width=3.6in]{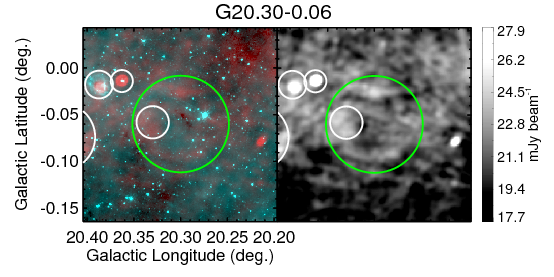}
\includegraphics[width=3.6in]{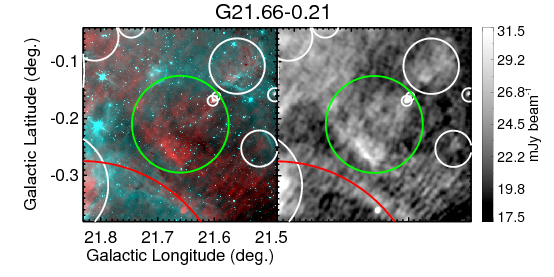}
\includegraphics[width=3.6in]{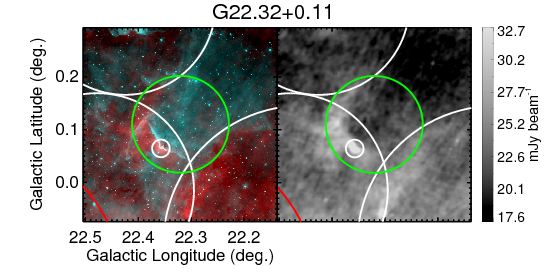}
\end{figure*}
\begin{figure*}
\includegraphics[width=3.6in]{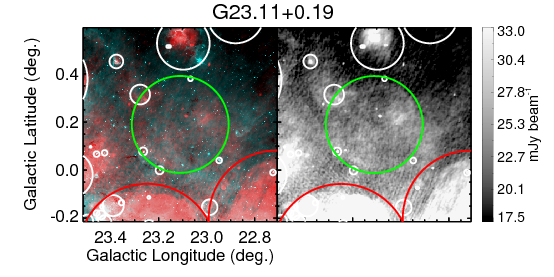}
\includegraphics[width=3.6in]{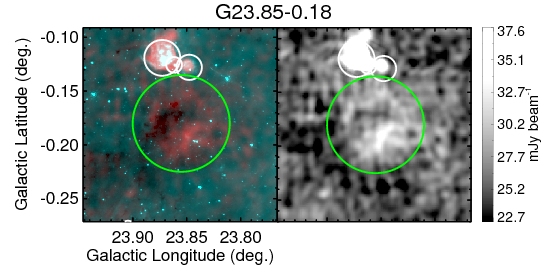}
\includegraphics[width=3.6in]{{G25.49+0.01_THOR_two}.png}
\includegraphics[width=3.6in]{{G26.04-0.42_THOR_two}.png}
\includegraphics[width=3.6in]{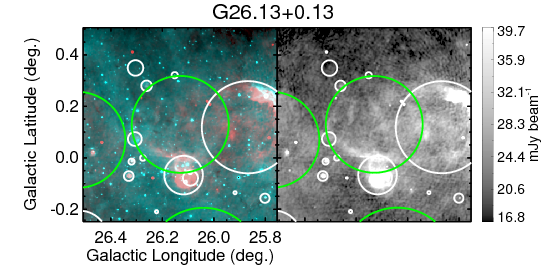}
\includegraphics[width=3.6in]{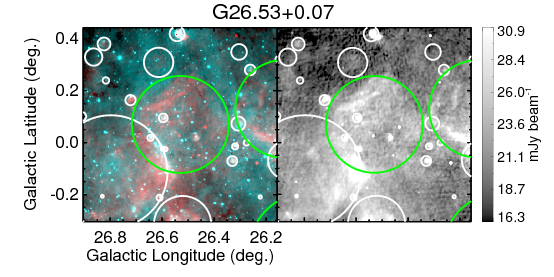}
\includegraphics[width=3.6in]{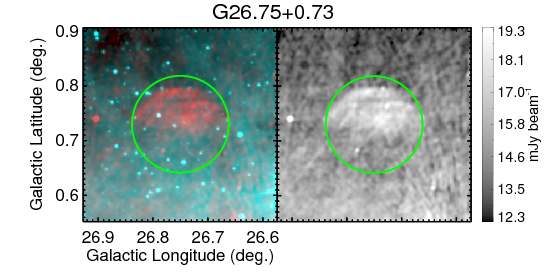}
\includegraphics[width=3.6in]{{G27.06+0.04_THOR_two}.png}
\includegraphics[width=3.6in]{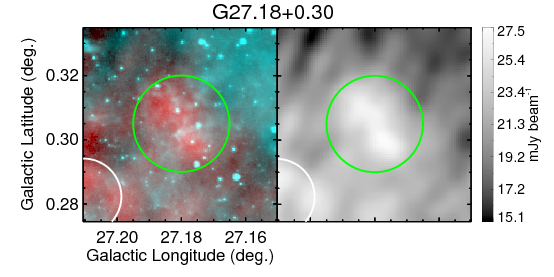}
\includegraphics[width=3.6in]{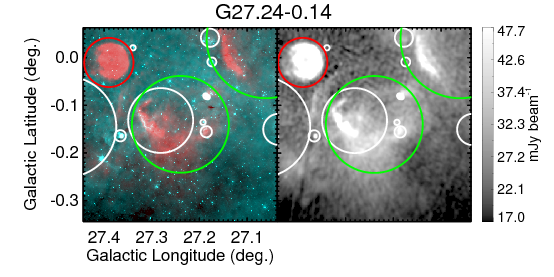}
\caption{Images for all identified SNR candidates.  The left panels show two-color GLIMPSE
       8.0\,\micron\ (cyan) and THOR+VGPS 21\,cm (red), as in
       Fig.~\ref{fig:snrs}.  The right panels show THOR+VGPS data
       alone.  Circles in both panels are the same as in
       Fig.~\ref{fig:snrs}, with candidate SNRs are enclosed by green
       circles, known SNRs by red circles, and known or candidate
       \hii\ regions by white circles.}
\end{figure*}
\begin{figure*}
\includegraphics[width=3.6in]{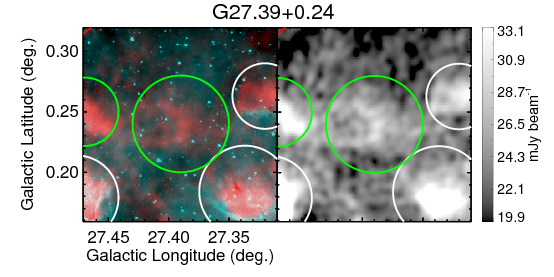}
\includegraphics[width=3.6in]{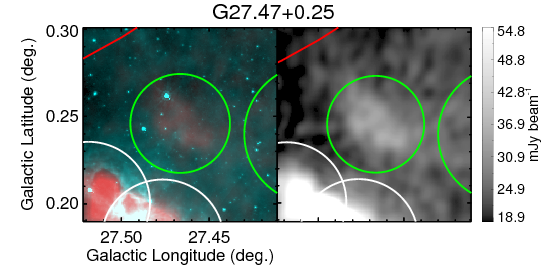}
\includegraphics[width=3.6in]{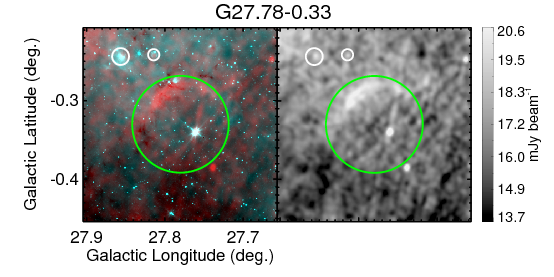}
\includegraphics[width=3.6in]{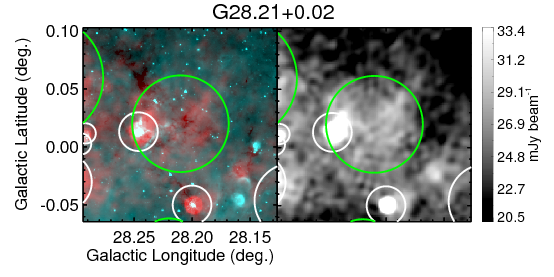}
\includegraphics[width=3.6in]{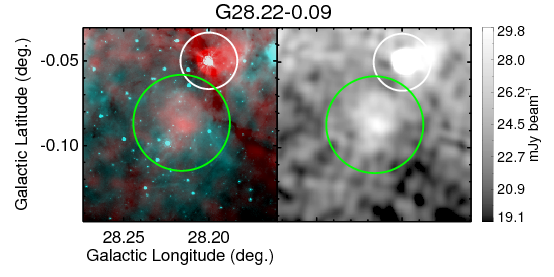}
\includegraphics[width=3.6in]{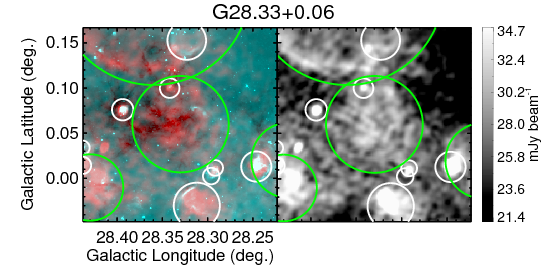}
\includegraphics[width=3.6in]{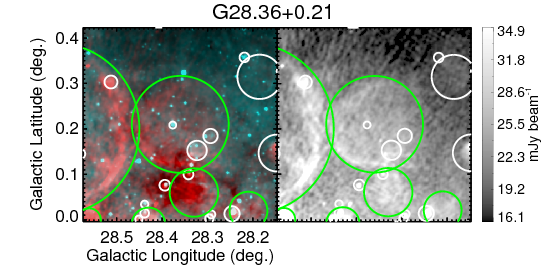}
\includegraphics[width=3.6in]{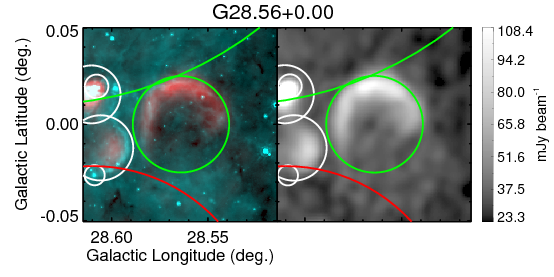}
\includegraphics[width=3.6in]{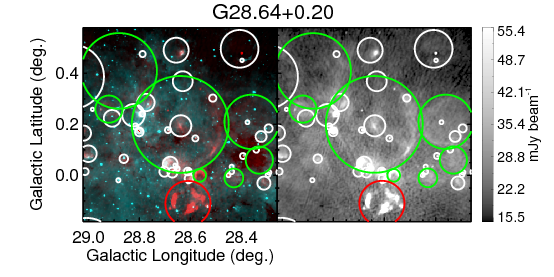}
\includegraphics[width=3.6in]{{G28.78-0.44_THOR_two}.png}
\end{figure*}
\begin{figure*}
\includegraphics[width=3.6in]{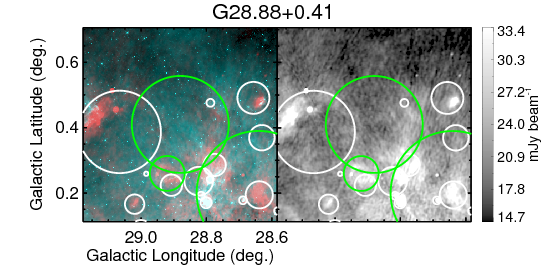}
\includegraphics[width=3.6in]{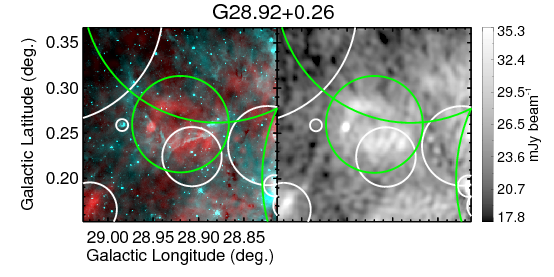}
\includegraphics[width=3.6in]{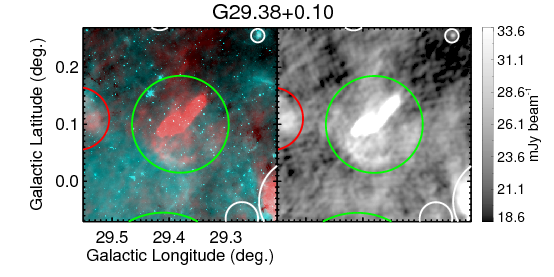}
\includegraphics[width=3.6in]{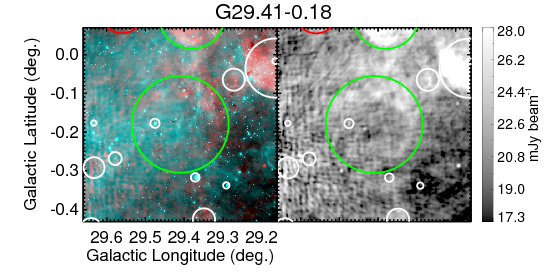}
\includegraphics[width=3.6in]{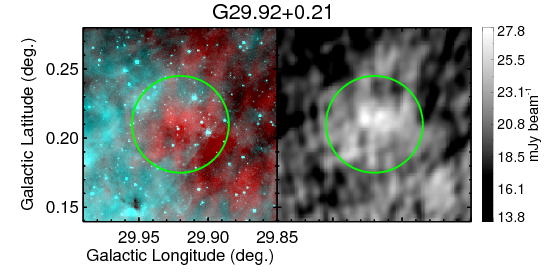}
\includegraphics[width=3.6in]{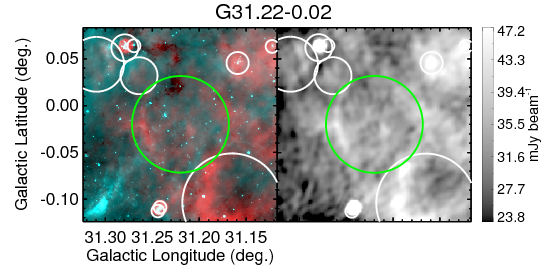}
\includegraphics[width=3.6in]{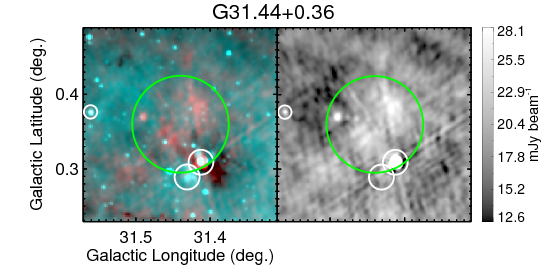}
\includegraphics[width=3.6in]{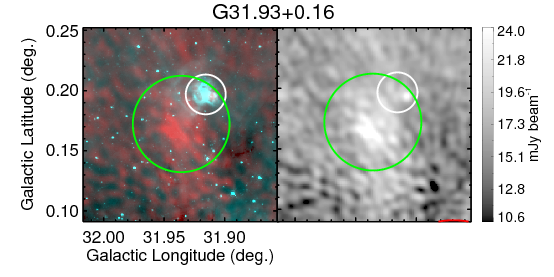}
\includegraphics[width=3.6in]{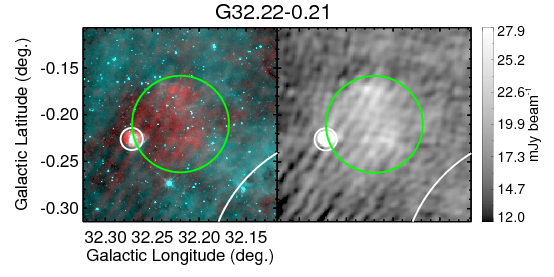}
\includegraphics[width=3.6in]{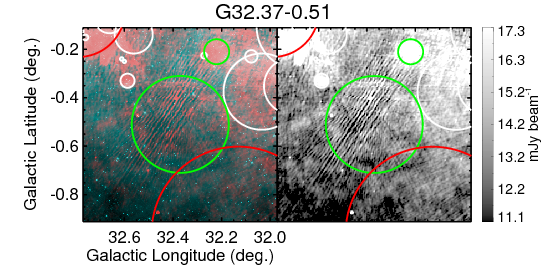}
\end{figure*}
\begin{figure*}
\includegraphics[width=3.6in]{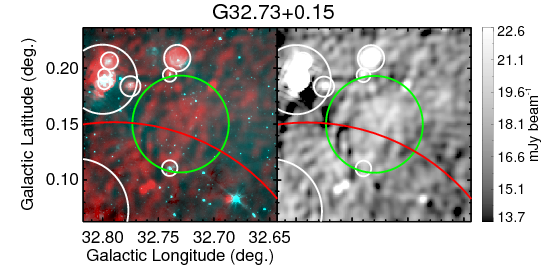}
\includegraphics[width=3.6in]{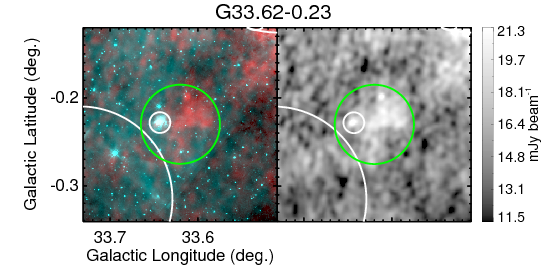}
\includegraphics[width=3.6in]{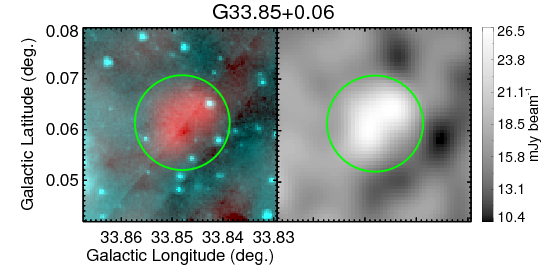}
\includegraphics[width=3.6in]{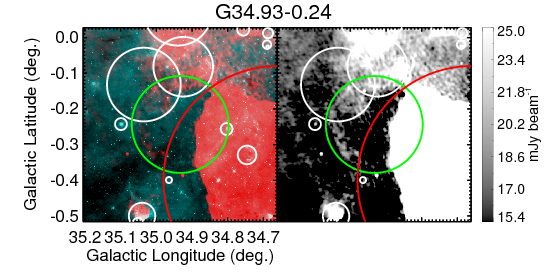}
\includegraphics[width=3.6in]{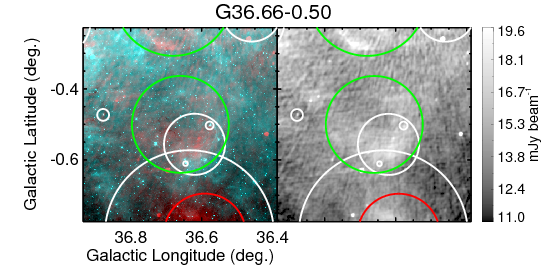}
\includegraphics[width=3.6in]{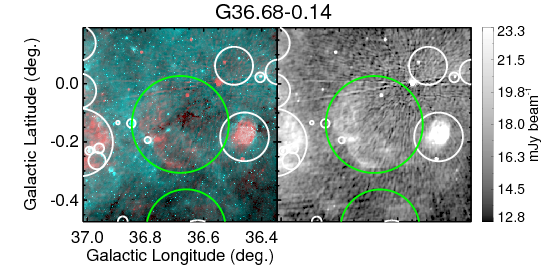}
\includegraphics[width=3.6in]{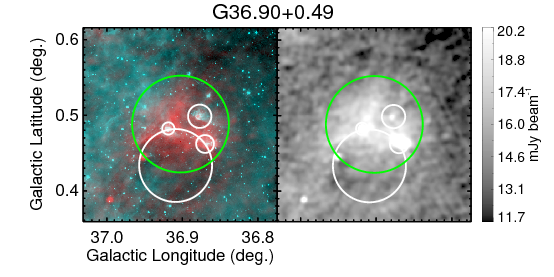}
\includegraphics[width=3.6in]{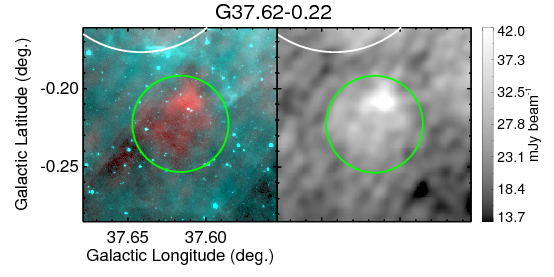}
\includegraphics[width=3.6in]{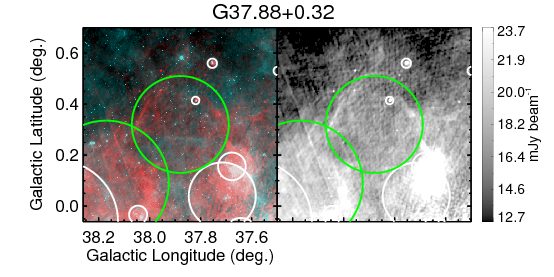}
\includegraphics[width=3.6in]{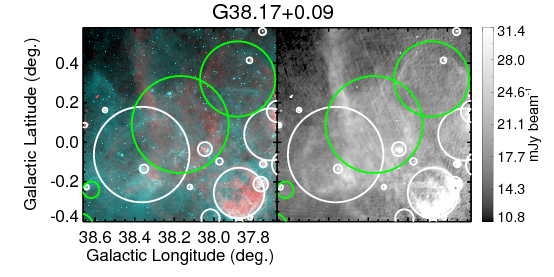}
\end{figure*}
\begin{figure*}
\includegraphics[width=3.6in]{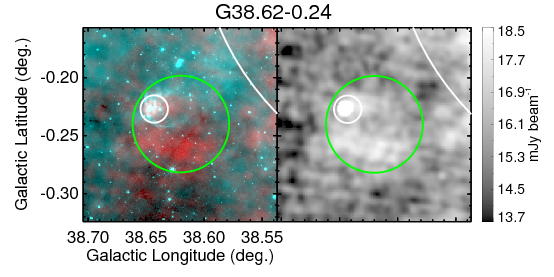}
\includegraphics[width=3.6in]{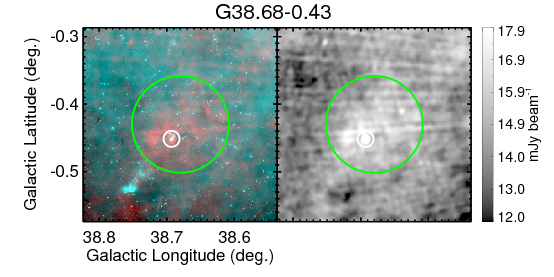}
\includegraphics[width=3.6in]{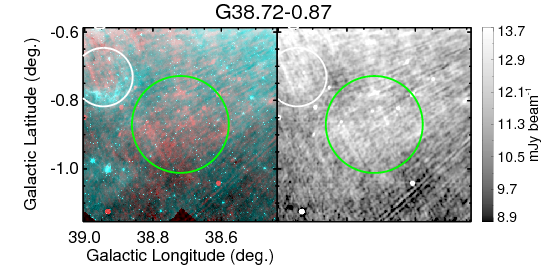}
\includegraphics[width=3.6in]{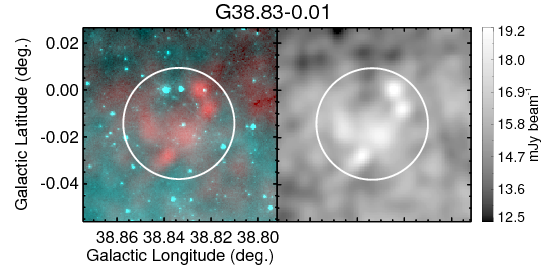}
\includegraphics[width=3.6in]{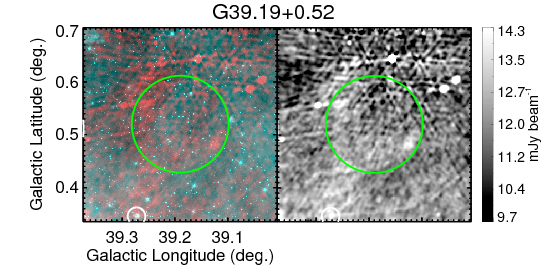}
\includegraphics[width=3.6in]{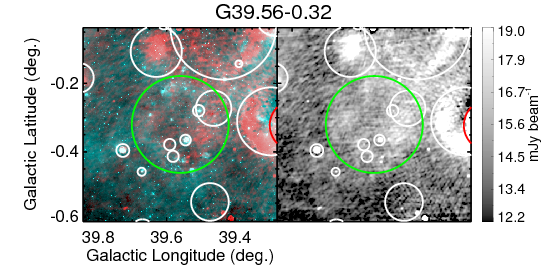}
\includegraphics[width=3.6in]{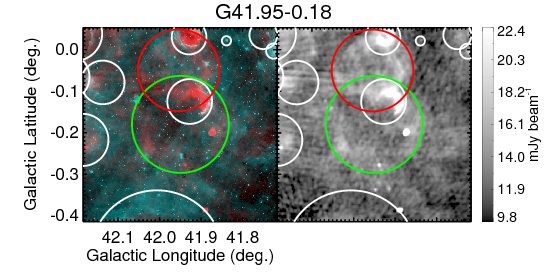}
\includegraphics[width=3.6in]{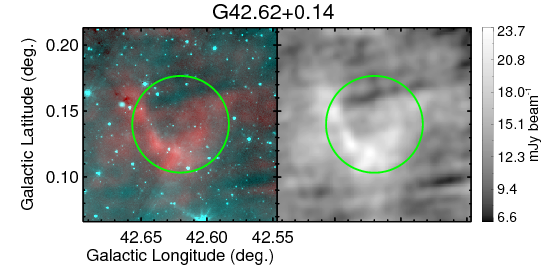}
\includegraphics[width=3.6in]{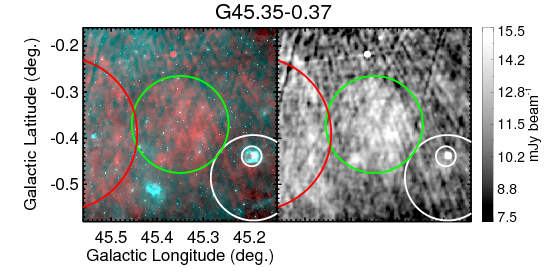}
\includegraphics[width=3.6in]{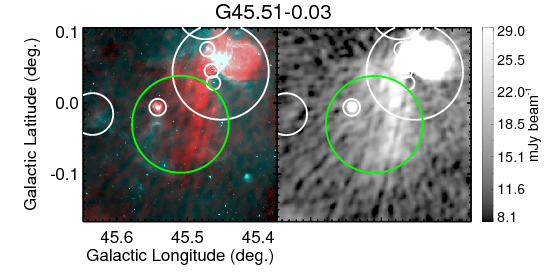}
\end{figure*}
\begin{figure*}
\includegraphics[width=3.6in]{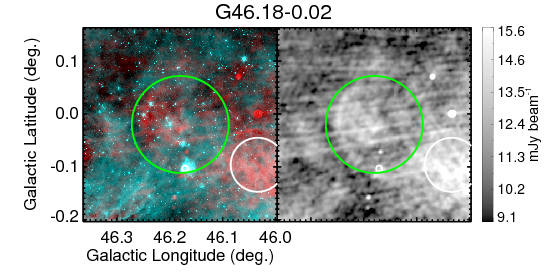}
\includegraphics[width=3.6in]{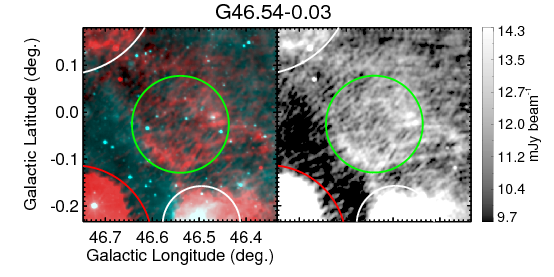}
\includegraphics[width=3.6in]{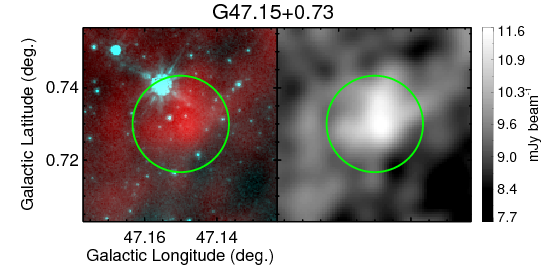}
\includegraphics[width=3.6in]{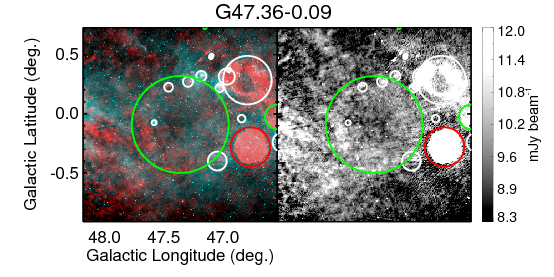}
\includegraphics[width=3.6in]{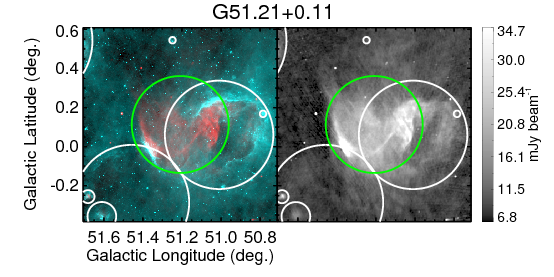}
\includegraphics[width=3.6in]{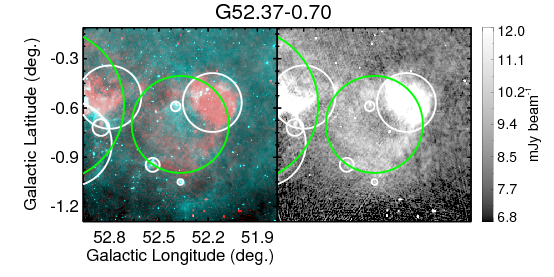}
\includegraphics[width=3.6in]{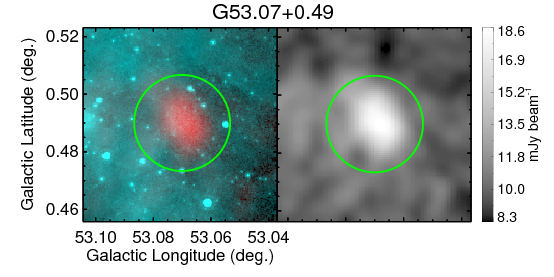}
\includegraphics[width=3.6in]{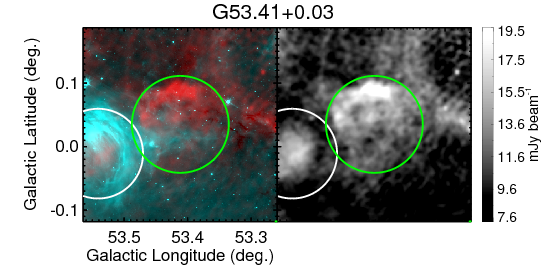}
\includegraphics[width=3.6in]{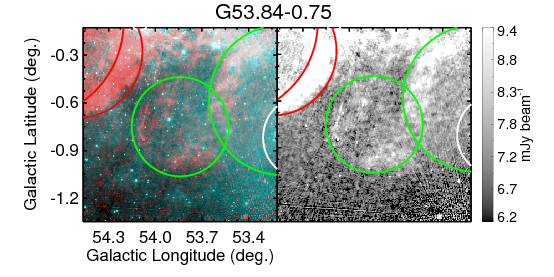}
\includegraphics[width=3.6in]{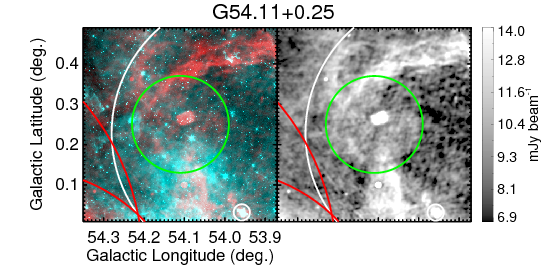}
\end{figure*}
\begin{figure*}
\includegraphics[width=3.6in]{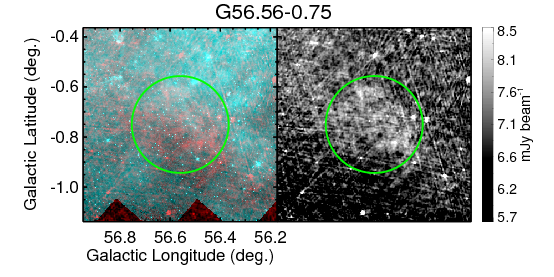}
\includegraphics[width=3.6in]{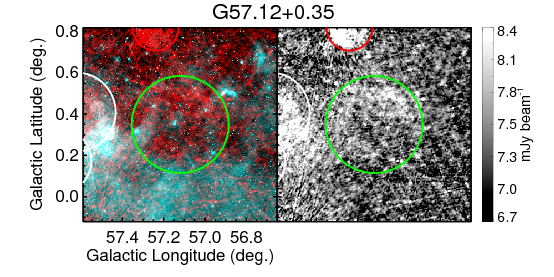}
\includegraphics[width=3.6in]{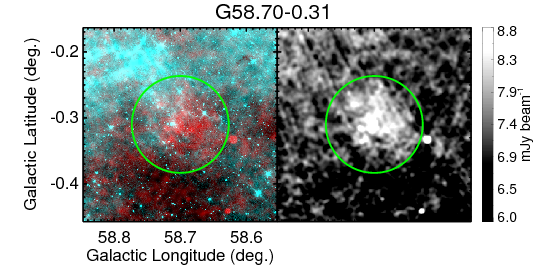}
\includegraphics[width=3.6in]{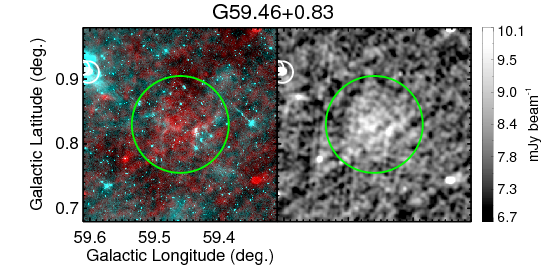}
\includegraphics[width=3.6in]{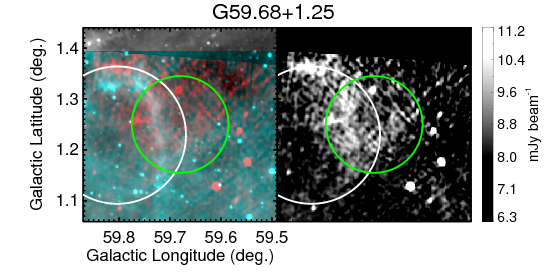}
\includegraphics[width=3.6in]{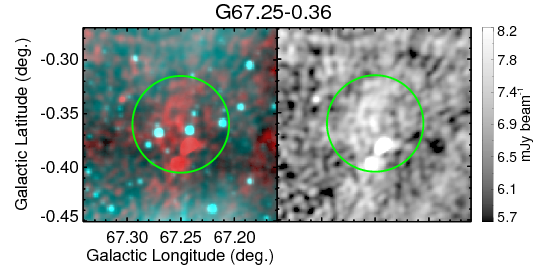}
\end{figure*}

\end{appendix}

\end{document}